\documentclass[preprint,10pt]{aastex}

\newcommand{\fin}{\phi_{\rm in}}

\usepackage{epsfig}
\usepackage{graphicx}
\usepackage{amsmath,amssymb}
\usepackage{natbib}
\usepackage{subfigure}

\begin{document}

\title{Strong field effects on pulsar arrival times: general orientations}
\author{Yan Wang}

\affil{Department of Astronomy, Nanjing University, Nanjing 210093,
  China, and Center for Gravitational Wave Astronomy and Department of
  Physics and Astronomy, University of Texas at Brownsville,
  Brownsville, Texas 78520}
\author{Teviet Creighton,  Richard H. Price and Frederick A.~Jenet}
\affil{Center for Gravitational Wave Astronomy and Department of
Physics and Astronomy, University of Texas at Brownsville,
Brownsville, Texas 78520}

\begin{abstract}

  A pulsar beam passing close to a black hole can provide a probe of
  very strong gravitational fields even if the pulsar itself is not in
  a strong field region. In the case that the spin of the hole can be
  ignored, we have previously shown that all strong field effects on
  the beam can be understood in terms of two ``universal'' functions,
  $F(\phi_{\rm in})$ and $T(\phi_{\rm in})$ of the angle of beam
  emission $\phi_{\rm in}$; these functions are universal in that they
  depend only on a single parameter, the pulsar/black hole distance
  from which the beam is emitted.  Here we apply this formalism to
  general pulsar-hole-observer geometries, with arbitrary alignment of
  the pulsar spin axis and arbitrary pulsar beam direction and angular
  width. We show that the analysis of the observational problem has
  two distinct elements: (i)~the computation of the location and
  trajectory of an observer-dependent ``keyhole'' direction of
  emission in which a signal can be received by the observer; (ii)~the
  determination of an annulus that represents the set of directions
  containing beam energy.  Examples of each are given along with an
  example of a specific observational scenario.
\end{abstract}
\maketitle

\section{Introduction}\label{sec:intro}

\subsection{Background}
There has been much recent interest in the possibility of pulsars near
the Galactic
center\citep{pfahlloeb04,freitagetal2006,lazioetal2006,munoetal2008}.
It is therefore possible that radio telescopes may receive pulsar
beams that have passed close to the supermassive black hole (SMBH) in
Sgr A*. In an earlier paper\citep{paperI}, hereafter ``Paper I,'' we
showed how such observations would encode information about the strong
field region near the SMBH. In a separate paper\citep{probpaper} we
discuss what observational program will be appropriate to search for
pulsar beams deflected by the SMBH in Sgr A$^*$.

In Paper I we simplified the analysis by assuming that the SMBH in Sgr
A* is not rotating. It is not plausible, of course, for the hole to be
strictly nonrotating, but it is likely that the rotation rate is of
order $J/M^2=a/M\sim0.5$\citep{meliabook}, rather than close to the
extremal limit $a/M=1$. (Here and throughout we adopt the conventions
that $G=c=1$.)  The most significant effect of rotation will be to
allow prograde photon orbits to come closer to the horizon without
being captured, and to require retrograde orbits to be further from
the horizon.  An estimate of the effect can be made relatively easily
for photon orbits in the equatorial plane of a Kerr hole, for photons
that are prograde (orbital angular momentum aligned with the hole's
spin angular momentum) and retrograde (angular momenta anti-aligned).
For emission of equatorial photons at a (Boyer-Lindquist) radial
coordinate $30M$ around  a hole with $a/M= 0.5$, the critical angle for 
a prograde photon is the same as if it were emitted at radius $38.4M$
near a Schwarzschild hole; for a retrograde photon the critical angle 
is the same as that for a photon emitted at radius $25.3M$ around 
a Schwarzschild hole. A more complete analysis of spin effects will be
deferred to a later paper; for now we note that while spin effects are
non-negligible, they will not be dominant.

While keeping in mind its limitations, we use
the nonrotating assumption due to its great simplification. As shown 
in Paper I, for a nonrotating hole all black hole effects on a pulsar beam 
are contained in two 
relatively simple functions, one of which, $F(\fin)$, describes the
bending of the pulsar beam due to gravitational effects, while the
other, $T(\fin)$, describes the gravitational and geometric time
delay. In these functions $\fin$ is the angle, with respect to the
direction radially away from the SMBH, at which the pulsar beam is emitted in
the astronomical (not the comoving) frame. These functions are
parameterized only by $r_0/M$ the Schwarzschild radial coordinate
location $r_0$ of the emission event in units of the geometrized mass
$M$. It is this simplification that distinguishes our approach from 
those of others who have studied the problem both with further approximations
and with more numerically intensive investigations 
\citep{campanaparodistella95,gorham86,goicoecheaetal92,
oscozetal97,lagunawolszcan97}.

\begin{figure}[h]
\begin{center}
\includegraphics[width=.8\textwidth]{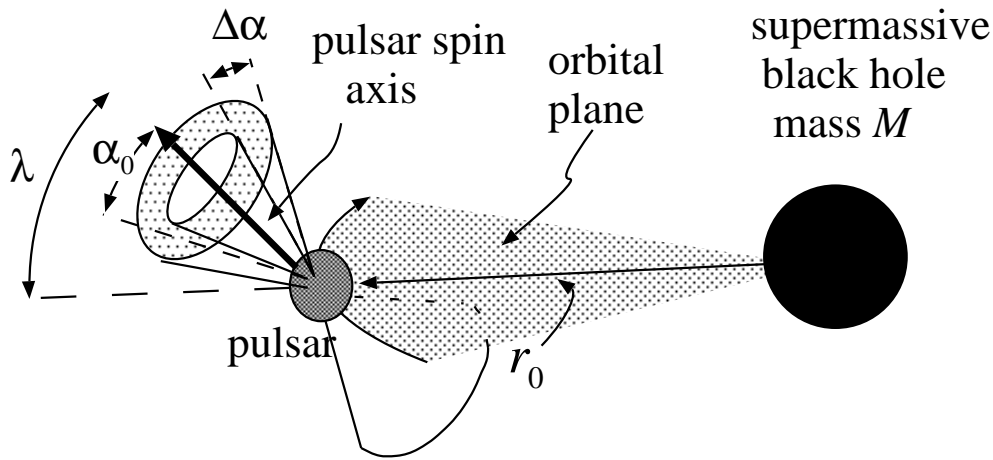}
\caption{Configuration of pulsar/SMBH system. } \label{fig:model}
\end{center}
\end{figure}

\subsection{Geometrical configuration}\label{subsec:config}

The basic geometry of our model is shown in Fig.~\ref{fig:model}.  At
the event of emission, the radial distance from the SMBH (as measured by the
Schwarzschild radial coordinate) is $r_0$, To describe the beam, we go
to the frame of the pulsar, the frame in which physical processes
produce the beam. 
In this frame the pulsar spin axis is at an angle
$\lambda$ above the orbital plane, and the angle $\alpha$ describes
the location of elements of the beam relative to the pulsar spin
axis. The rapidly rotating beam is confined to a conical region
extending around the spin axis from $\alpha=\alpha_0-\Delta\alpha/2$
to $\alpha=\alpha_0+\Delta\alpha/2$. We will use the term ``beam
cone'' to refer to this region.

We note at the outset
that it is often useful to view the problem of beam propagaton from
the changing location of (but not in the frame of) the pulsar. That
is, at any time we will see things from the point of view of an
observer at rest in the astronomical rest frame at the location of the
pulsar. For this observer, the distant stars and the distant Earth
will remain in fixed angular positions, but the SMBH will move in 
an orbit (not necessarily a circular, or even closed orbit).

\subsection{Outline}\label{subsec:outline}

Sections \ref{sec:keyholes} and \ref{sec:annulus} give a general
framework for understanding the geometry of strong field
effects. These two sections separate the problem into two parts.  The
first part, that in Sec.~\ref{sec:keyholes}, deals with the
geometrical relationship of the pulsar location, SMBH location
and Earth location. For any such relationship we establish a
``keyhole'' direction, a direction in which radiation from the pulsar,
i.e., part of the pulsar beam, will be gravitationally bent so that it
is directed to the Earth.  The second part of the geometry, that of
Sec.~\ref{sec:annulus}, deals with the directions of pulsar
beaming. Due to the aberration of photon directions (the ``headlight
effect'') these directions depend not only on the configuration
parameters $\lambda$, $\alpha_0$, $\Delta\alpha$, but also on the
velocity of the pulsar motion.  For any model, and any pulsar
velocity, this gives us an annulus of beaming. The beaming will reach
the Earth if the keyhole position falls within the beaming annulus.
In Sec.~\ref{sec:scenarios} we put together the techniques of the
previous two sections to show the observational results for a
particular astrophysical model.
A summary and conclusions are given in Sec.~\ref{sec:conc}.

\section{Pulsar-SMBH-Earth geometry and the keyhole location}\label{sec:keyholes}

For a given location of the Earth, at any moment in the orbit of the
pulsar there is a given direction for a photon that will follow an
SMBH-deflected orbit and end up at the Earth. We call this the
{\em keyhole} direction.  This direction is found using the
universal function $F(\fin)$ of Paper I.  If the keyhole direction
overlaps the beam cone in which the pulsar is actually sending energy,
then there will be a pulse sent off towards the Earth roughly once per
rotation. The precise times of pulse emission and reception are then
found from the assumptions about the pulsar, and from $T(\fin)$, the
second universal function in Paper I. The issues of overlap with the
beam cone, and of timing, will be taken up in the subsequent two
sections. Here we focus only on the calculation of the keyhole
direction.

\subsection{Coordinate system}

We use polar coordinates with the orbital plane of the pulsar-SMBH
system defining the equator of the coordinate system, and with
directions specified by longitude and latitude coordinates
$(\varphi,\lambda)$.  In this coordinate system the Earth has some
fixed location $(\lambda_\oplus,\varphi_\oplus)$.
We define our coordinate system to be a global inertial coordinate system
whose origin is instantaneously centered on the pulsar at the
time of emission of any given photon.  Thus every photon trajectory
starts at the origin, and the SMBH position is fixed for any photon
trajectory, but the black hole moves around the origin as a function
of emission time.

\begin{figure}[h]
\begin{center}
\includegraphics[width=.4\textwidth]{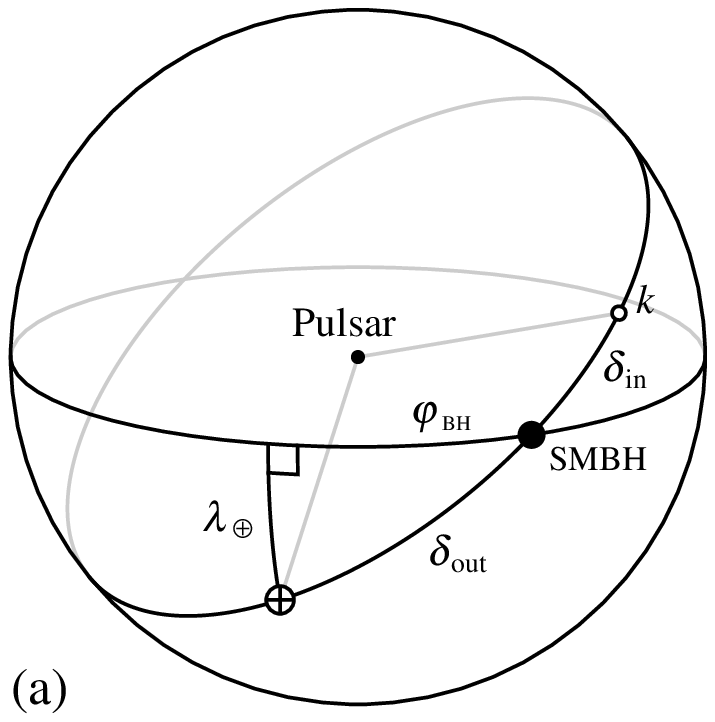}\quad\quad
\includegraphics[width=.4\textwidth]{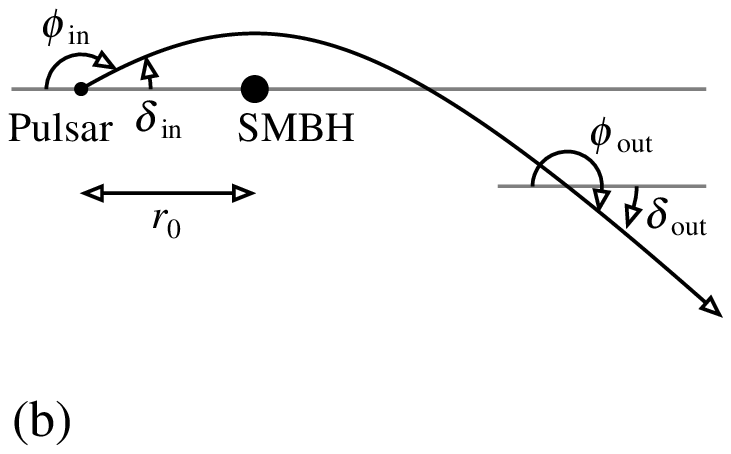}
\caption{Photon initial and outgoing directions in spherical coordinates and within the trajectory plane. } \label{fig:trajectory}
\end{center}
\end{figure}

At any time the configuration of the system can be defined by the
longitude $\varphi_\mathrm{BH}$ of the SMBH (relative to the Earth)
and the latitude $\lambda_\oplus$ of the Earth (relative to the
orbital plane).  Any photon traveling from the pulsar to the Earth
must do so on a plane containing the Earth, the pulsar, and the SMBH.
This defines a great circle in our coordinate system, with an angular
separation $\delta_\mathrm{out}$ between the Earth and the SMBH, given by
$\cos\delta_\mathrm{out}=\cos\lambda_\oplus\cos(\varphi_\mathrm{BH}-\varphi_\oplus)$,
 as shown in
Figure~\ref{fig:trajectory}(a).  To find the keyhole direction, we
must find the corresponding angle $\delta_\mathrm{in}$, that is, the
initial photon direction relative to the pulsar-SMBH axis.

Figure~\ref{fig:trajectory}(b) shows the photon trajectory within the
trajectory plane, labeling the initial and outgoing photon directions
$\phi_\mathrm{in}$ and $\phi_\mathrm{out}$ according to the
conventions in Paper I.  Given $\delta_\mathrm{out}$ above, we
find $\phi_\mathrm{out}=\pi-\delta_\mathrm{out}$.  We can then compute
$\phi_\mathrm{in}$ by numerically inverting the first universal
function $\phi_\mathrm{in}=F^{-1}(\phi_\mathrm{out};r_0)$ for the
pulsar-SMBH separation $r_0$ at the time of emission.  We
next compute $\delta_\mathrm{in}=\pi-\phi_\mathrm{in}$.  Lastly, the longitude
$\varphi_k$ and latitude $\lambda_k$ of the keyhole are
obtained by spherical geometry.

If a pulsar beam is significantly deflected, it will be spread out 
in the plane of the photon trajectory. As discussed in Paper I, this
leads to a change in the received beam strength  by
the intensity amplification factor 
\begin{equation}
  \label{Amp}
  \mbox{Amp}=\frac{\sin\fin}{\sin{(F(\fin))}\,dF/d\fin}\,.
\end{equation}
This amplification factor depends only on $\fin$, and hence only on the pulsar-SMBH-Earth
configuration because it omits the red/blueshift effects 
due to the pulsar motion. Those effects, in any case, are usually very small 
compared to geometrical
effects.

It should be noted that the keyhole direction is not unique;
adding  multiples
$2n\pi$ to $\phi_\mathrm{out}$, with integer $n$, gives the same outgoing direction, but
inverting the universal function yields a different initial angle
$\phi_\mathrm{in}$.  These correspond to photon trajectories that
orbit the SMBH one or more times in a prograde or retrograde
sense before emerging in the desired direction.  In general, $n=0$
corresponds to the ``direct'' (least-bent) path. For $\phi_{\rm out}>0$
the  $n=-1$ beam is
 the most
significant of the more strongly-bent paths (passing around the SMBH
 in the opposite sense to the direct path), and $n=+1$ is the path that
completes one full orbit in the same sense as the direct path, and so
on.

\subsection{Keyhole locations}

Our final algorithm for computing the keyhole direction
$(\varphi_k,\lambda_k)$ for a given SMBH longitude
$\varphi_\mathrm{BH}$ and Earth longitude and latitude
$(\varphi_\oplus,\lambda_\oplus)$ can be reduced to the following
equations,
\begin{eqnarray}
\phi_\mathrm{out} &=& \arccos\left(-\cos\lambda_\oplus
\cos(\varphi_\mathrm{BH}-\varphi_\oplus)\right)
\quad\in\quad[0,\pi) \label{algorithm1}\\
\phi_\mathrm{in} &=& F^{-1}\left(\phi_\mathrm{out}+2n\pi\;;\;r_0\right) \label{algorithm2} \\
\varphi_k &=& \varphi_\mathrm{BH} +
\arctan\!2\left(-\sin\phi_\mathrm{in}\cos\lambda_\oplus
\sin(\varphi_\mathrm{BH}-\varphi_\oplus)\;,\;
-\cos\phi_\mathrm{in}\sqrt{1-\cos^2\lambda_\oplus
  \cos^2(\varphi_\mathrm{BH}-\varphi_\oplus)}\right)\label{algorithm3} \\
\lambda_k &=& \arcsin\left(\sin\fin\,\sin\lambda_\oplus
/\sqrt{1-\cos^2(\varphi_\mathrm{BH}-\varphi_\oplus)\cos^2\lambda_\oplus}\right)
\quad\in\quad[-\pi/2,\pi/2]\,,\label{algorithm4}
\end{eqnarray}
where $\arctan\!2(y,x)$ is (as in some programming languages) the
inverse trig function returning the argument of $x+iy$ in the correct
quadrant.

\begin{figure}[h]
\begin{center}
\includegraphics[width=.45\textwidth]{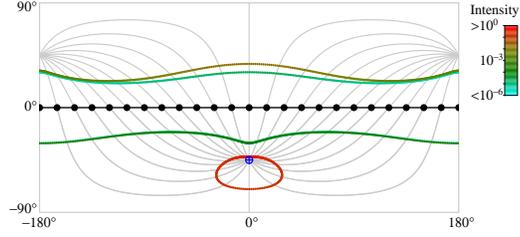}
\caption{Keyhole positions and attenuation factors for $r_0=10M$, for various looping numbers.} \label{fig:windings}
\end{center}
\end{figure}

Figure~\ref{fig:windings} gives a map of keyhole positions for Earth
location $\lambda_\oplus=-\pi/4$ (shown as a blue $\oplus$), for all
possible SMBH longitudes $\varphi_\mathrm{BH}-\varphi_\oplus$,
assuming $r_0=10GM/c^2$.  Since each keyhole position corresponds to a
specific $\phi_\mathrm{in}$, we also compute the intensity attenuation
or amplification given in Eq.~\eqref{Amp}, and color the keyhole point
accordingly.  This lets us see immediately which photon paths will be
interesting (i.e.\ not too attenuated by bending).  Also plotted are
gray curves of the great circles corresponding to the trajectory
plane, shown for every $15^\circ$ in
$\varphi_\mathrm{BH}-\varphi_\oplus$.  The way to read the graph is as
follows: For any given SMBH position, find the gray curve connecting
it to the Earth.  This is the trajectory plane of photons that can
reach the Earth for that SMBH position.  Find where that curve
intersects one of the colored keyhole curves.  That intersection
indicates the keyhole location: the initial direction in which photons
must be beamed to reach the Earth.  The color indicates the relative
geometric attenuation of photons traveling along that path.  We can
clearly see the direct paths (the red loop), the most gently bent
indirect paths (brown line), and more highly-wound paths: there is a
countably infinite number of such possible highly-wound paths, but the
amplification factor become exponentially small with increasing windings.

\begin{figure}[h]
\begin{center}
\includegraphics[width=.45\textwidth]{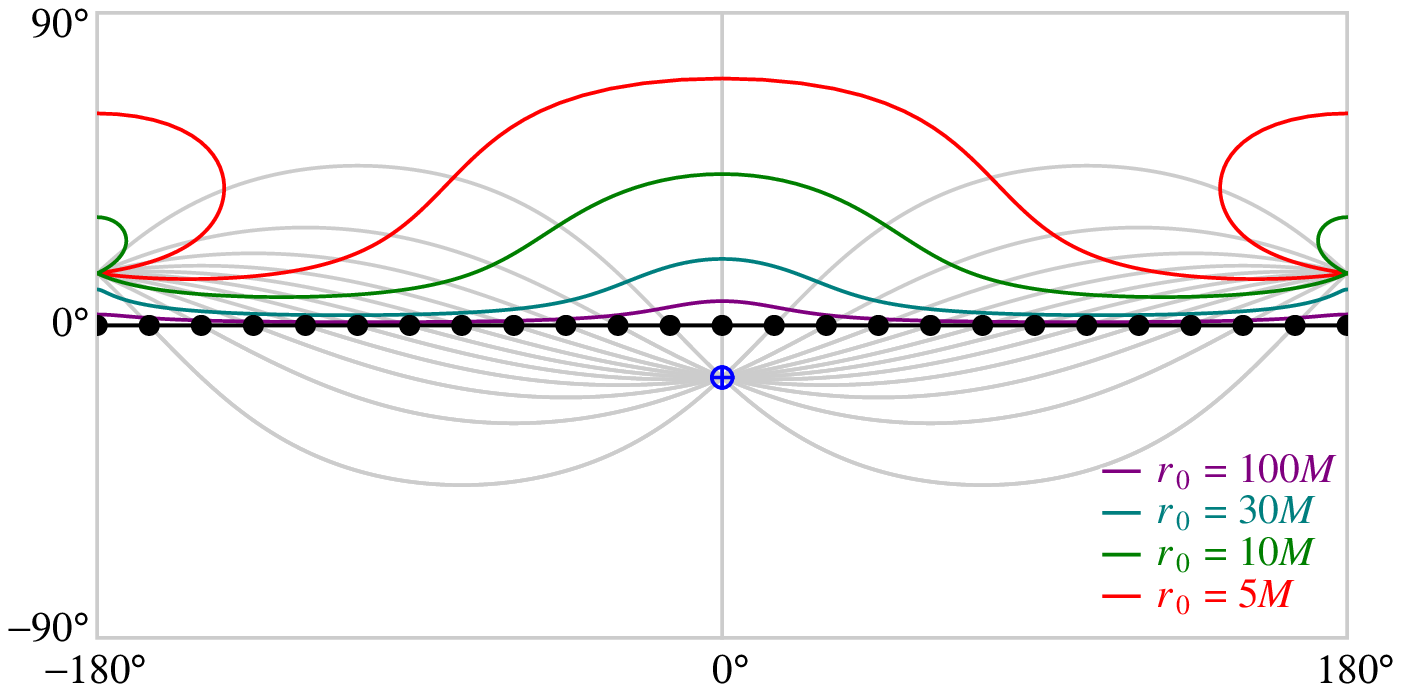}\quad
\includegraphics[width=.45\textwidth]{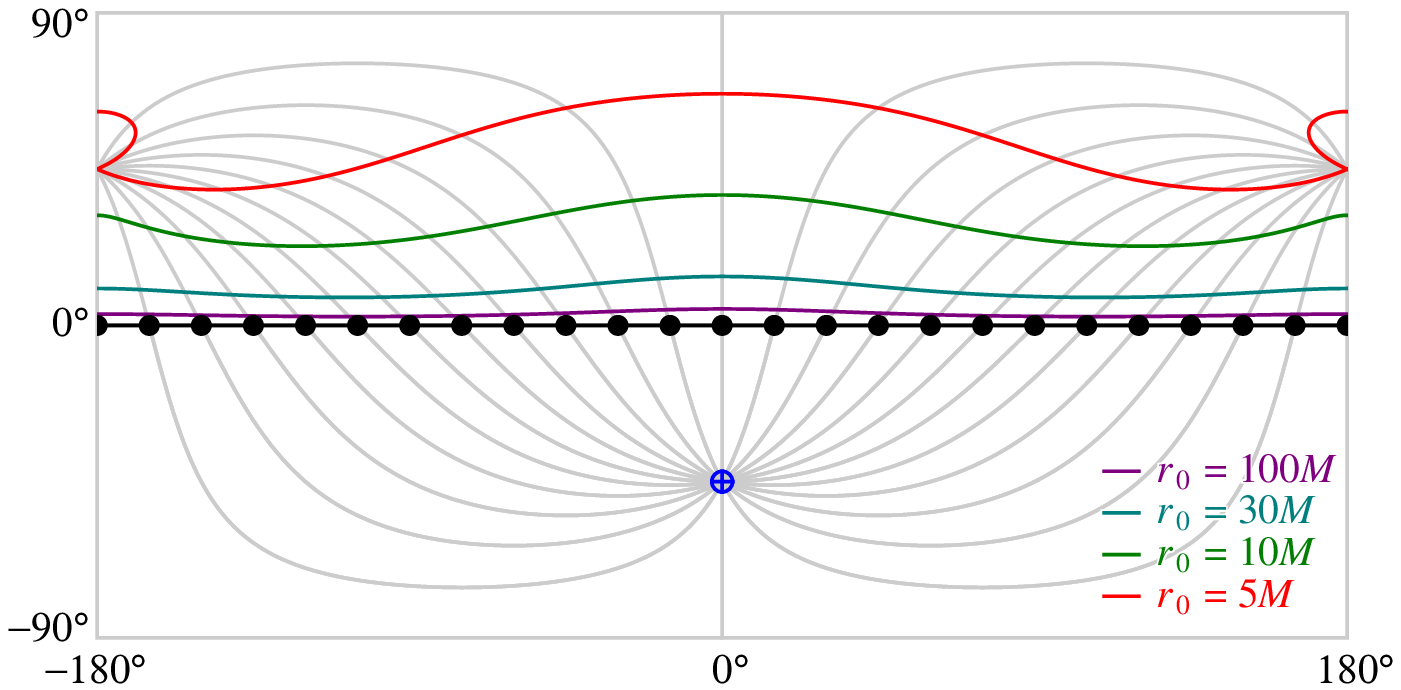}\\[2ex]
\includegraphics[width=.45\textwidth]{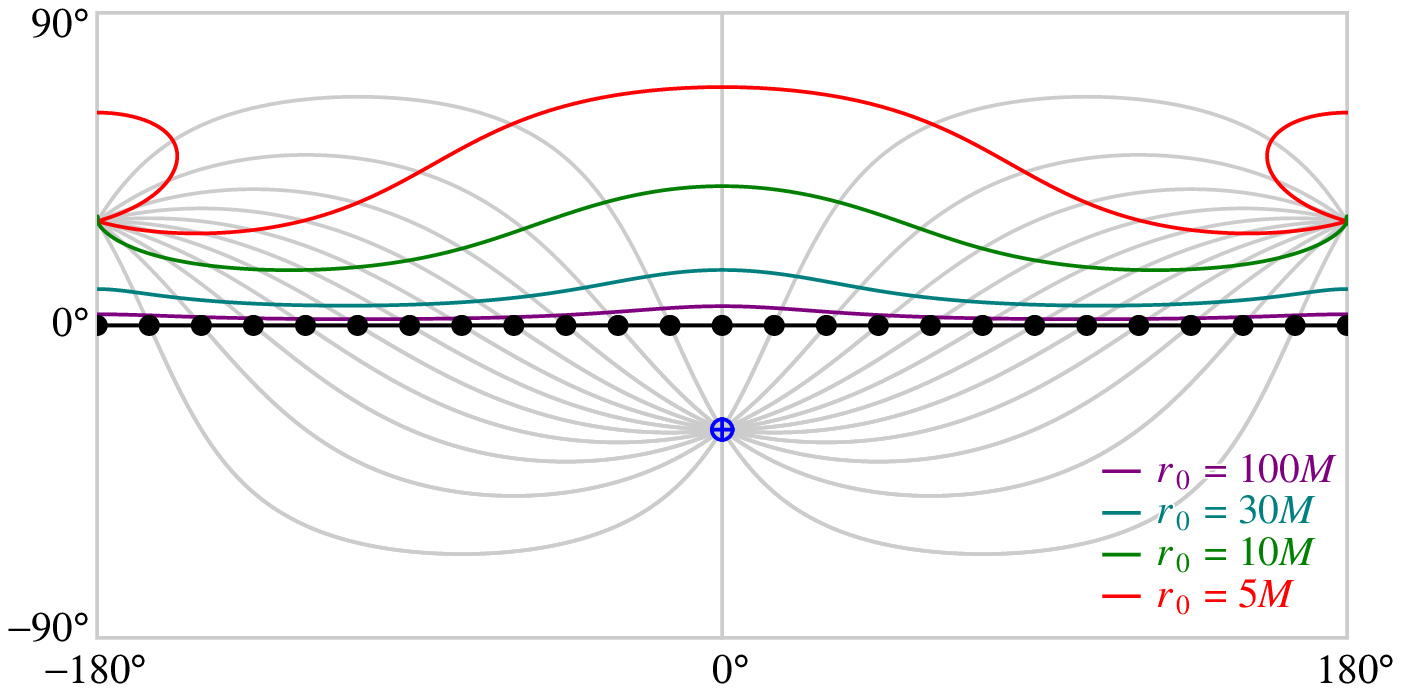}\quad
\includegraphics[width=.45\textwidth]{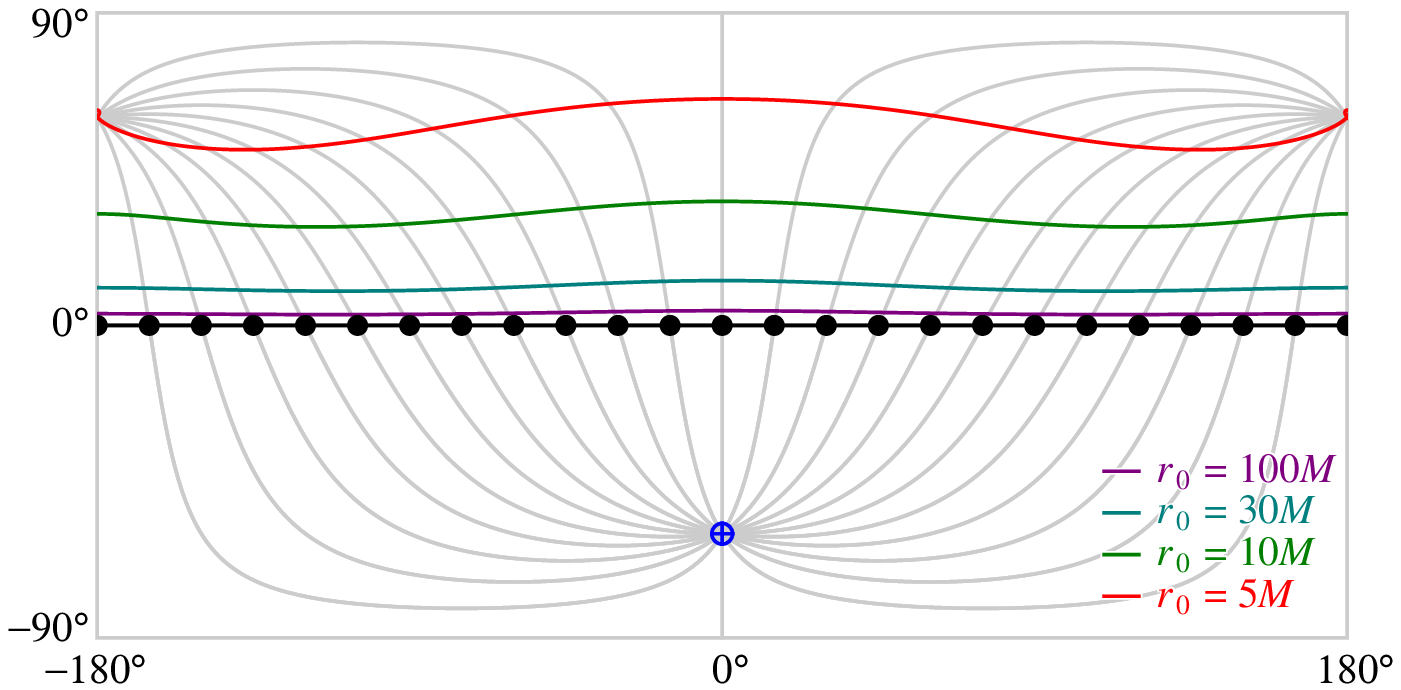}
\caption{Keyhole positions for 
$\lambda_\oplus=-15^\circ$, $-30^\circ$, $-45^\circ$, and $-60^\circ$ for various $r_0$.} \label{fig:keyholes}
\end{center}
\end{figure}

\begin{figure}[h]
\begin{center}
\includegraphics[width=.45\textwidth]{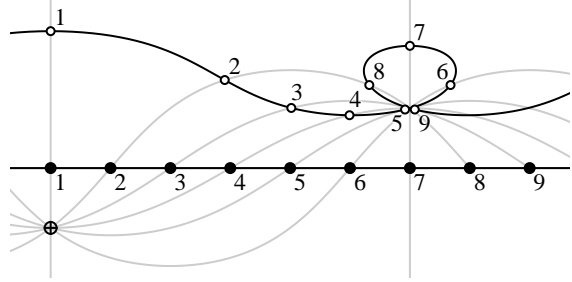}\quad
\caption{Detail of SMBH and corresponding keyhole positions for $\lambda_\oplus=-30^\circ$, $r_0=5GM/c^2$} \label{fig:keyholezoom}
\end{center}
\end{figure}

In Figure~\ref{fig:keyholes} we focus our attention on the ``$n=-1$''
trajectories, the most significant and gently bent of the
``indirect'' beam paths.  In this case, colors denote the pulsar-SMBH
distance $r_0$, as indicated.  Keyhole maps are shown for
$\lambda_\oplus=-15^\circ$, $-30^\circ$, $-45^\circ$, and $-60^\circ$.
Note that for sufficiently small $r_0$ and $\lambda_\oplus$ the simple
one-to-one relationship between SMBH and keyhole positions in
Figure~\ref{fig:windings} breaks down.  This occurs when there is a
SMBH position (relative to the pulsar) such that a beam aimed directly
away from the Earth is bent around to point to the Earth.  In such a
situation, due to symmetry about the Earth-pulsar line, there will in
fact be two SMBH positions (relative to the pulsar) that will give the
same deflection.  From the pulsar's perspective, as the SMBH orbits
it, the keyhole describes a loop in the sky starting from the
anti-Earth direction.  Figure~\ref{fig:keyholezoom} illustrates in
detail how to relate SMBH and keyhole positions for such a loop.


\newpage

\newpage

\section{The distorted emission annulus}\label{sec:annulus}

\begin{figure}[h]
\begin{center}
\includegraphics[width=.6\textwidth]{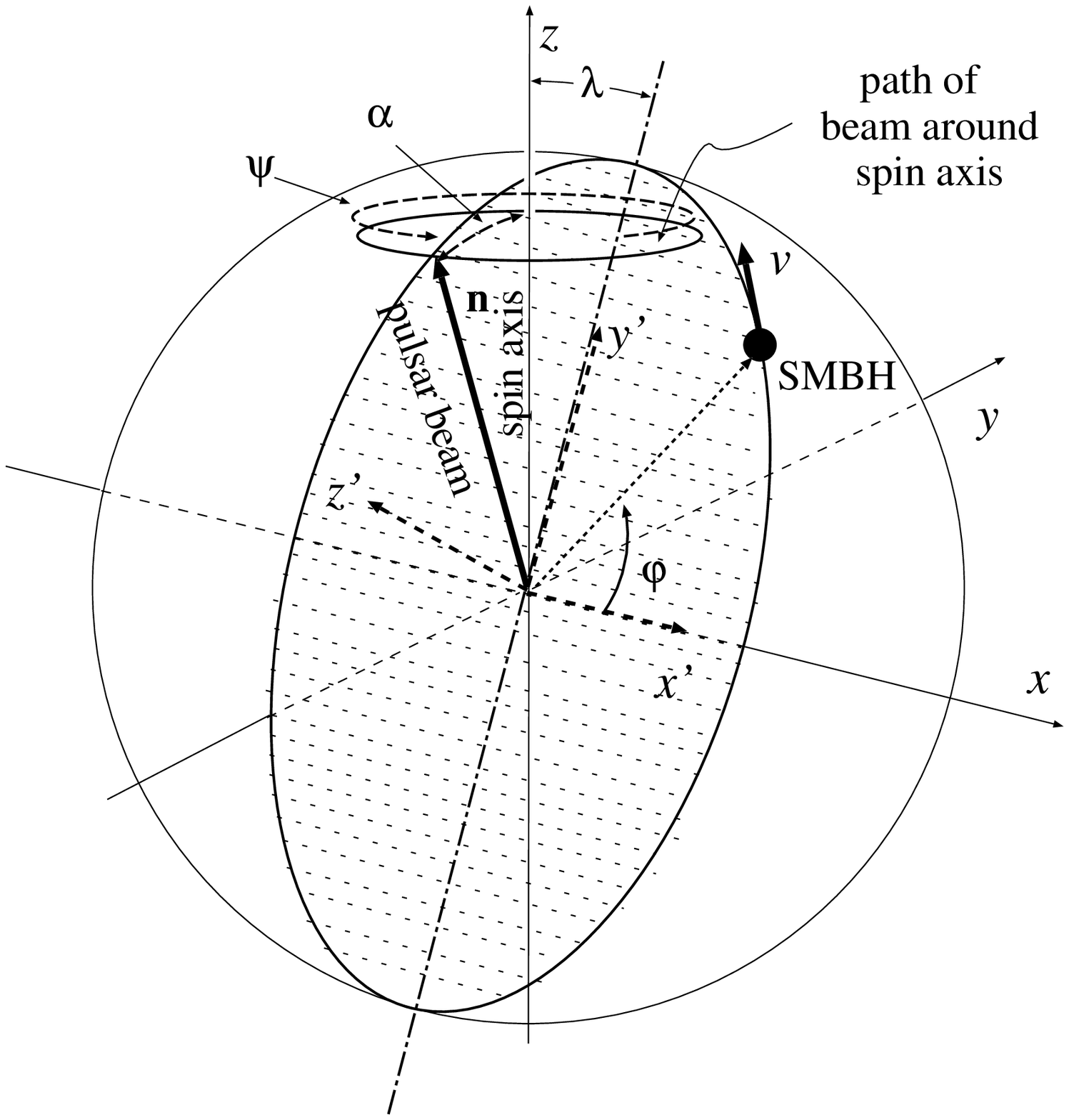}
\caption{Geometry for computation of the distorted annulus.} \label{fig:sphere2}
\end{center}
\end{figure}

In our model system, the pulsar is assumed to emit its beam in a conical region of angular
width $\Delta\alpha$ as shown in Fig.~\ref{fig:model}. That
description of emission directions applies in the frame
instantaneously comoving with the pulsar. To compute the influence of
strong field effects we must find the location of this region in the
``astronomical frame,'' the frame in which the SMBH is at
rest. It is in this astronomical frame that we can infer the angle $\phi_{\rm in}$
that is the input to the universal functions $F(\phi_{\rm in})$ and
$T(\phi_{\rm in})$.

The relationship between the emission direction in the pulsar frame
and the emission direction in the astronomical frame involves a
rotational transformation and a Lorentz boost. It therefore depends on
the geometric parameters $\alpha$ and $\lambda$ as pictured in
Fig.~\ref{fig:sphere2}. It also depends on the velocity with which the
pulsar moves with respect to the astronomical frame, or equivalently
on $-{\mathbf v}$, where ${\mathbf v}$ is the velocity of the SMBH
relative to the pulsar, as pictured in
Fig.~\ref{fig:sphere2}. Calculation of the aberration and rotation do
not need to assume anything about the pulsar orbit (or about the SMBH
orbit from the viewpoint of the pulsar), but it can be intuitively useful to
consider the SMBH to be in a circular orbit around the pulsar, with
radius $r_0$ and at azimuthal position $\varphi$, as indicated 
in Fig.~\ref{fig:sphere2}.  The translation
back and forth between the specification of ${\mathbf v}$ and of
the pair of parameters $r_0/M,\;\varphi$ is simple and straightforward.
What should be noted in particular is that -- unlike the keyhole
direction -- the aberration-distorted annulus of beam directions is
independent of the direction to the Earth observer.

To compute the distortion, we
use the general configuration of the pulsar and the SMBH
system and the  coordinate frames in Fig.~\ref{fig:sphere2}. The
$x$ axis coincides with the $x'$ axis. The $x'y'$ plane is the orbital plane
and is rotated about the $x$ axis by $\pi/2-\lambda$ from the plane perpendicular
to the pulsar spin axis.

We start in the unprimed system, the system aligned with the pulsar
spin axis, and in which the pulsar is at rest. In this system we
introduce the parameter $\psi$ to describe the phase of beam rotation,
so that at any one instant the direction of the beam is specified by
the unit vector ${\mathbf n}$ with components
$\{n_x,n_y,n_z\}=\{\sin\alpha\cos\psi, \sin\alpha\sin\psi,
\cos\alpha\}$. We find the components $n_{j'}$ in the $x'y'z'$
coordinate basis using $n_{j'}=T_{j'i}n_{i}$, where
\begin{equation}
  \label{Tij}
T_{j'i}=\left( \begin{array}{ccc}
1 & 0 & 0 \\
0 & \sin\lambda & \cos\lambda \\
0 & -\cos\lambda & \sin\lambda
\end{array} \right)\,,
\end{equation}
from which  we get
\begin{equation}\label{nprimes}
  \begin{split}
    n_{x'}&=\sin\alpha\cos\psi\\
n_{y'}&=\sin\lambda\sin\alpha\sin\psi+\cos\lambda\cos\alpha\\
n_{z'}&=\sin\lambda\cos\alpha-\cos\lambda\sin\alpha\sin\psi\,.
  \end{split}
\end{equation}

For a photon of energy  $E$, moving in the ${\mathbf n}$ direction, the 
components of the  4-momentum in the
$x'y'z'$ frame are
\begin{equation}\label{pmuprime}
p^{\mu'}=E\{1, n_{x'}, n_{y'}, n_{z'}\}\,.  
\end{equation}
 If the velocity of 
the SMBH in the pulsar rest frame has components $v_{x'},v_{y'}, v_{z'}$
then the 4-momentum components  $p^{\mu''}$ in the astronomical frame, the frame 
in which the SMBH
is stationary, are given by $p^{\mu''}=\Lambda^{\mu''}_{\ \ \nu'}\,p^{\nu'}$ where, in general,
\begin{equation}\label{genLorentz}
\Lambda^{\nu''}\,_{\!\mu'} = \left( \begin{array}{cc}
\gamma & -\gamma v_{i'} \\
-\gamma v_{j'} &
\delta_{i'j'}+v_{i'} v_{j'}(\gamma-1)/v^{2}
\end{array} \right)\,.
  \end{equation}
  If we assume a circular orbit for the black hole, with speed $v$,
  then $\{v_{x'},v_{y'},v_{z'}\}=v\{-\sin\varphi,\,\cos\varphi\,,0\}$.
  With these velocity components the matrix of Eq.~\eqref{genLorentz} becomes
\begin{equation}
\left( \begin{array}{cccc}
\gamma & \gamma\upsilon\sin\varphi & -\gamma\upsilon\cos\varphi & 0\\
\gamma\upsilon\sin\varphi &
1+(\gamma-1)\sin^{2}\varphi & -(\gamma-1)\sin\varphi\cos\varphi & 0\\
-\gamma\upsilon\cos\varphi &
-(\gamma-1)\sin\varphi\cos\varphi & 1+(\gamma-1)\cos^{2}\varphi & 0\\
0 & 0 & 0 & 1\\
\end{array} \right)\ .
\end{equation}
Multiplying this matrix
with the $p^{\mu'}$ components in Eq.~\eqref{pmuprime}, and using the values of
$n_{j'}$ given in Eqs.~\eqref{nprimes} gives us, finally, 
that the energy
$E''$ and direction cosines $n_{x''},n_{y''},n_{z''}$ are related to the
parameters $E,\alpha,\lambda,v,\psi$ in the pulsar emission frame by the following:
\begin{subequations}\label{boost}
\begin{align}
(E''/E)&=\gamma+\gamma\upsilon(\sin\alpha\cos\psi\sin\varphi-\sin\lambda\sin\alpha\sin\psi\cos\varphi
-\cos\lambda\cos\alpha\cos\varphi)\label{first}\\
n_{x''}(E''/E)&=\gamma\upsilon\sin\varphi+\sin\alpha\cos\psi(\gamma\sin^{2}\varphi+\cos^{2}\varphi)
-(\gamma-1)(\sin\lambda\sin\alpha\sin\psi+\cos\lambda\cos\alpha)\sin\varphi\cos\varphi\label{second}\\
n_{y''}(E''/E)&=-\gamma\upsilon\cos\varphi+(\sin\lambda\sin\alpha\sin\psi+\cos\lambda\cos\alpha)(\gamma\cos^{2}\varphi+\sin^{2}\varphi)-(\gamma-1)\sin\alpha\cos\psi\sin\varphi\cos\varphi
\label{third}\\
n_{z''}(E''/E)&=\sin\lambda\cos\alpha-\cos\lambda\sin\alpha\cos\psi\label{fourth}\,.
\end{align}
\end{subequations}
The factor $E''/E$ is the red/blueshift factor by which the photon
energy is modified, and $\gamma\equiv1/\sqrt{1-v^2}$ is the usual
Lorentz factor. The time for $\psi$ to go from 0 to $2\pi$ (i.e., the
pulsar spin period) is many orders of magnitudes smaller than the time
for $\varphi$ to go from 0 to $2\pi$ (i.e., the orbital time). We can
therefore take $\varphi$ to be constant while the pulsar beam traces
out a closed path. In this way
we get a closed path for any set of parameters
$\alpha,\lambda,{v}$.

\begin{figure}
\centering
\begin{tabular}{cc}
\epsfig{file=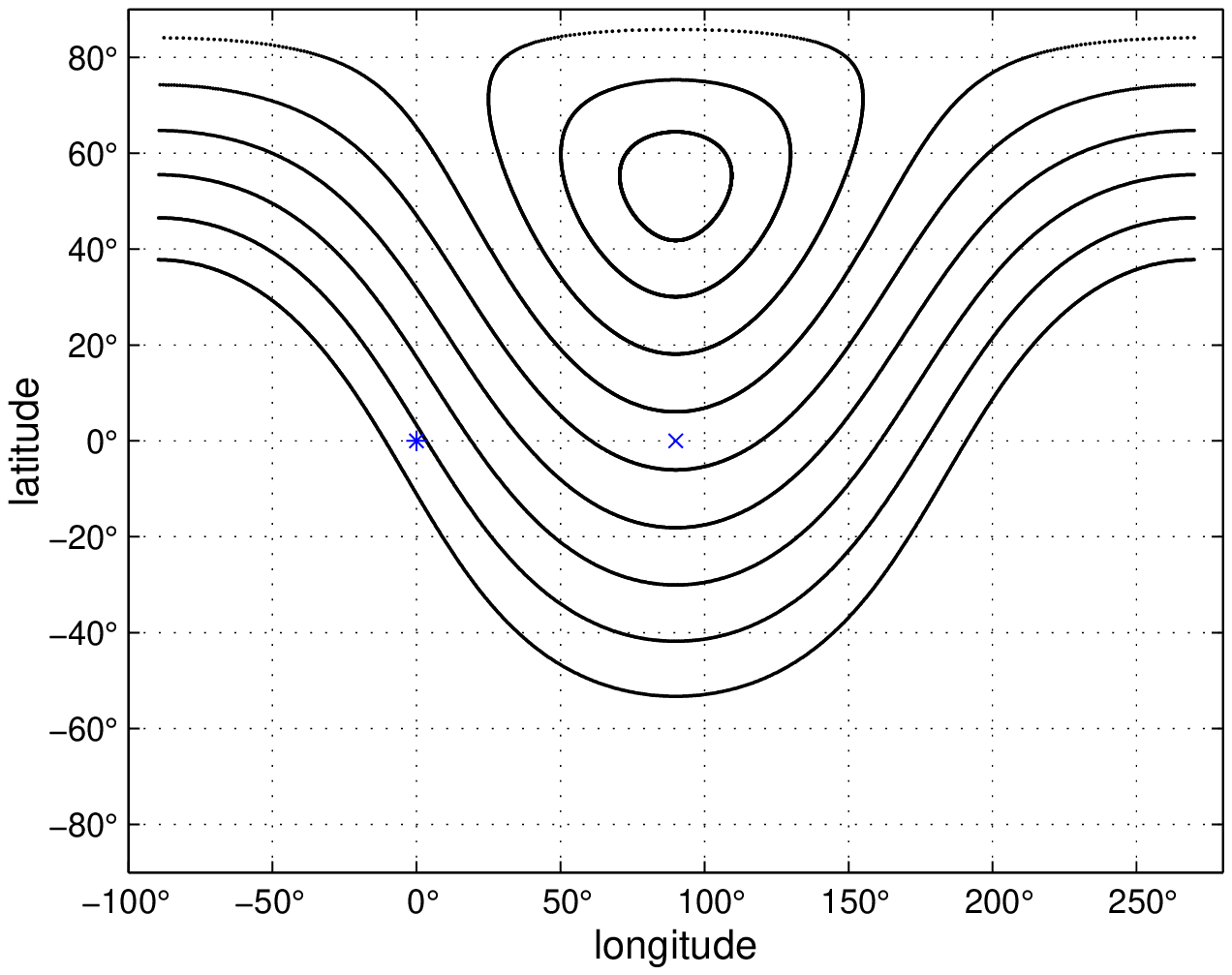,width=0.5\linewidth,clip=} &
\epsfig{file=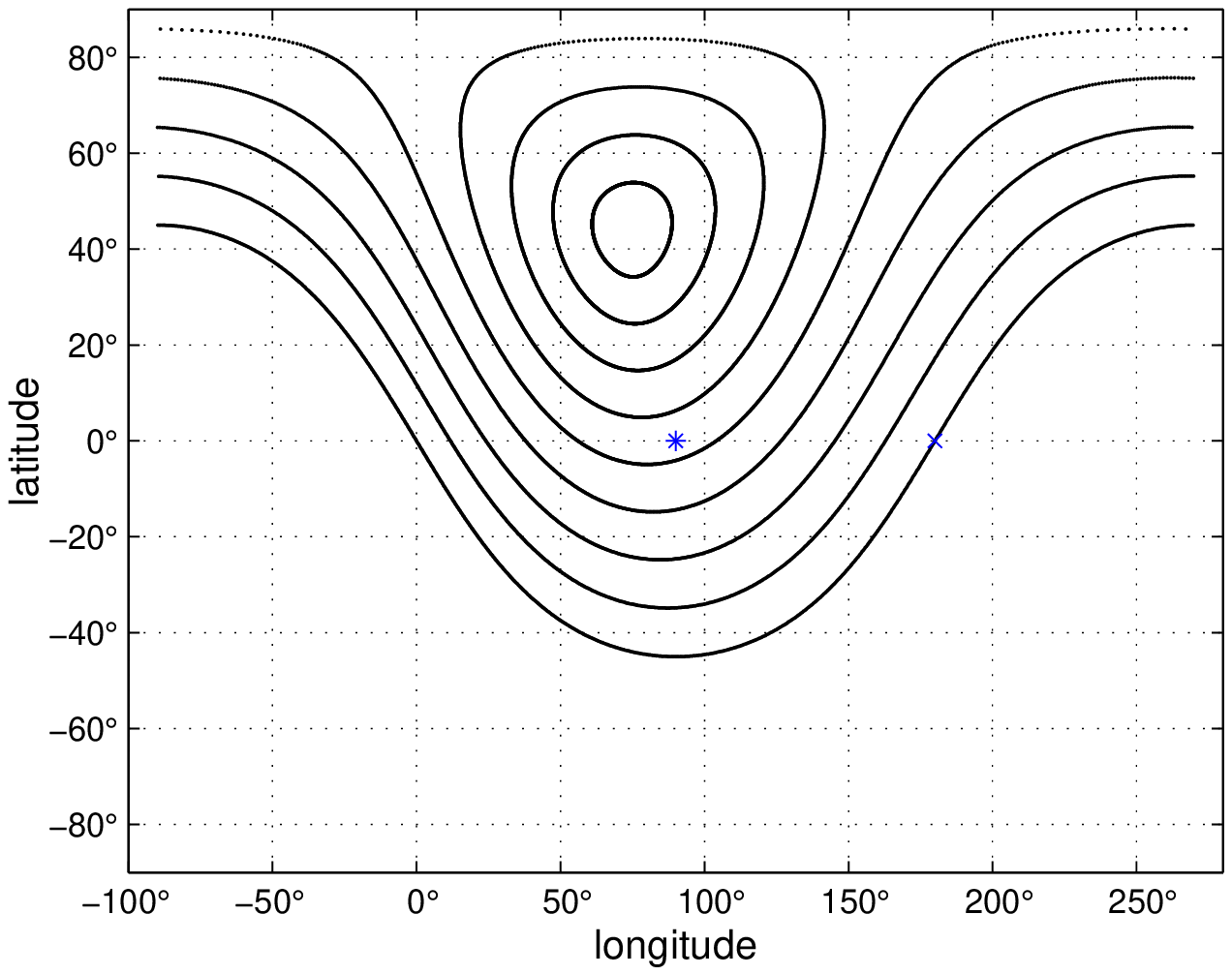,width=0.5\linewidth,clip=} \\
\epsfig{file=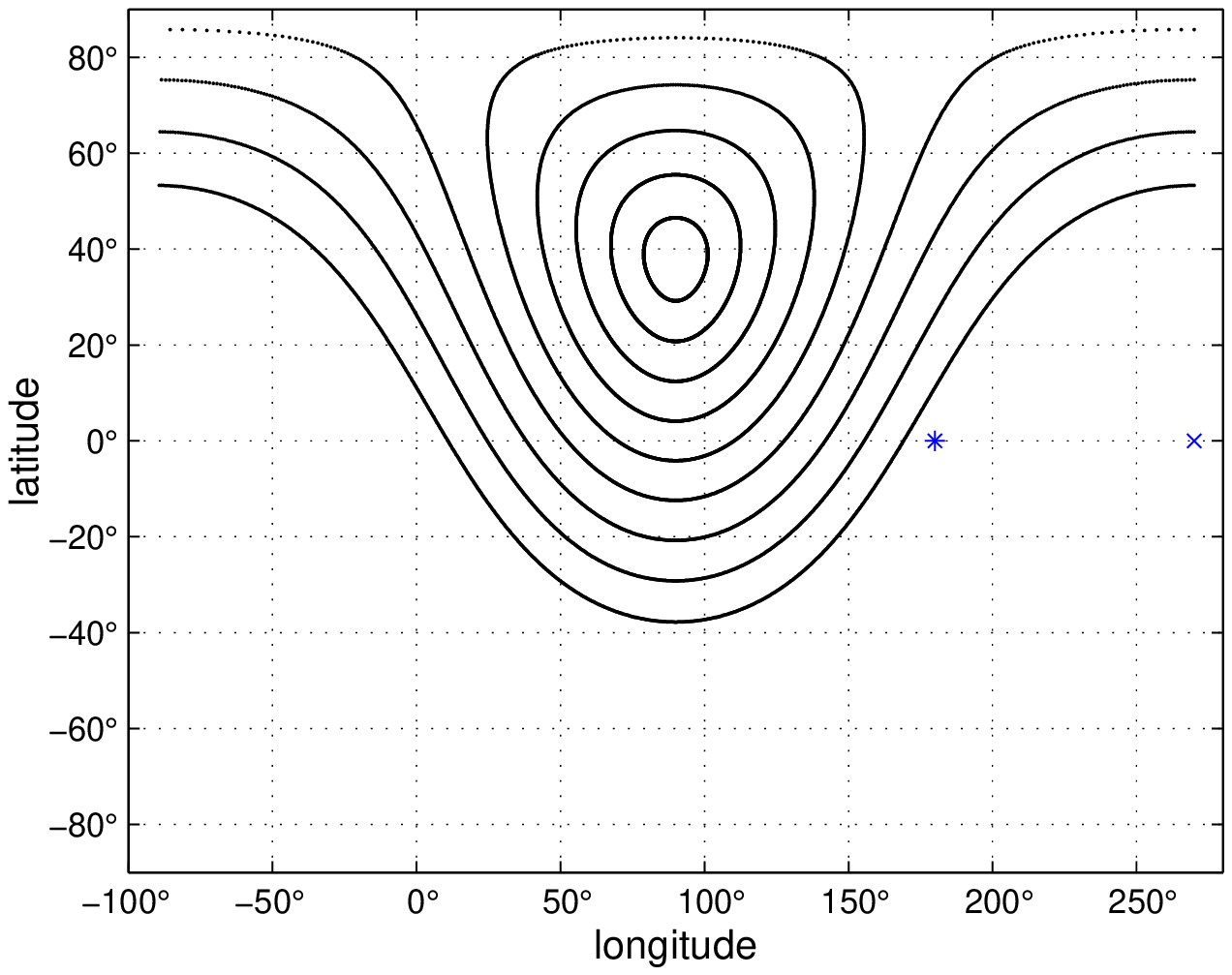,width=0.5\linewidth,clip=} &
\epsfig{file=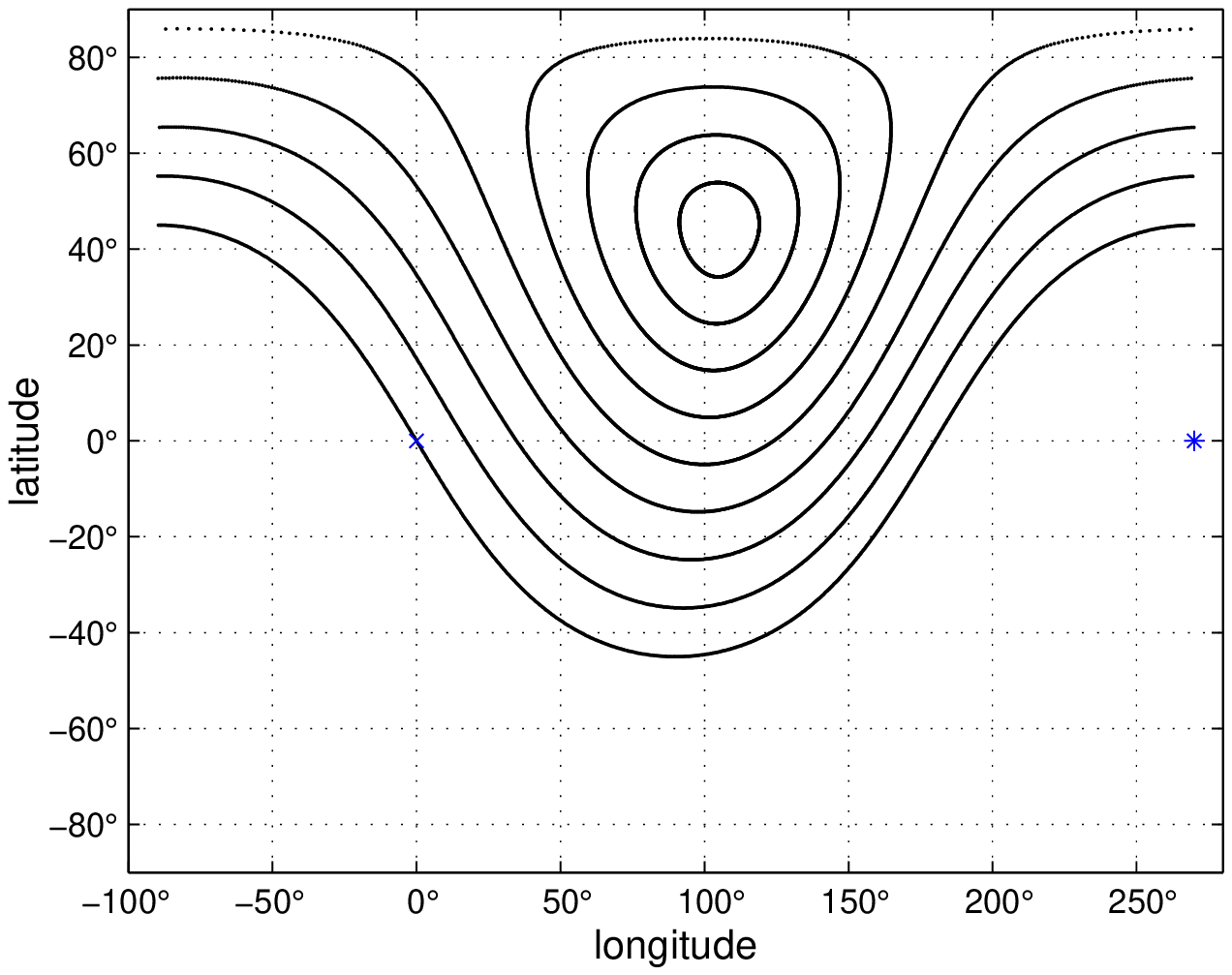,width=0.5\linewidth,clip=}
\end{tabular}
\caption{The distorted emission annulus for $r_{0}=30M$ and
$\lambda=45^{\circ}$. Each panel corresponds to a black hole
position in Fig.~\ref{fig:sphere2} at $\varphi=0^{\circ}$ (upper
  left); $\varphi=90^{\circ}$ (upper right); $\varphi=180^{\circ}$ (lower
  left); $\varphi=270^{\circ}$ (lower right). In each panel, an asterisk at
zero latitude marks the SMBH position and a cross marks the direction
of the velocity of the SMBH relative to the pulsar.} \label{fig:aberration1}
\end{figure}

\begin{figure}
\begin{center}
\begin{tabular}{cc}
\epsfig{file=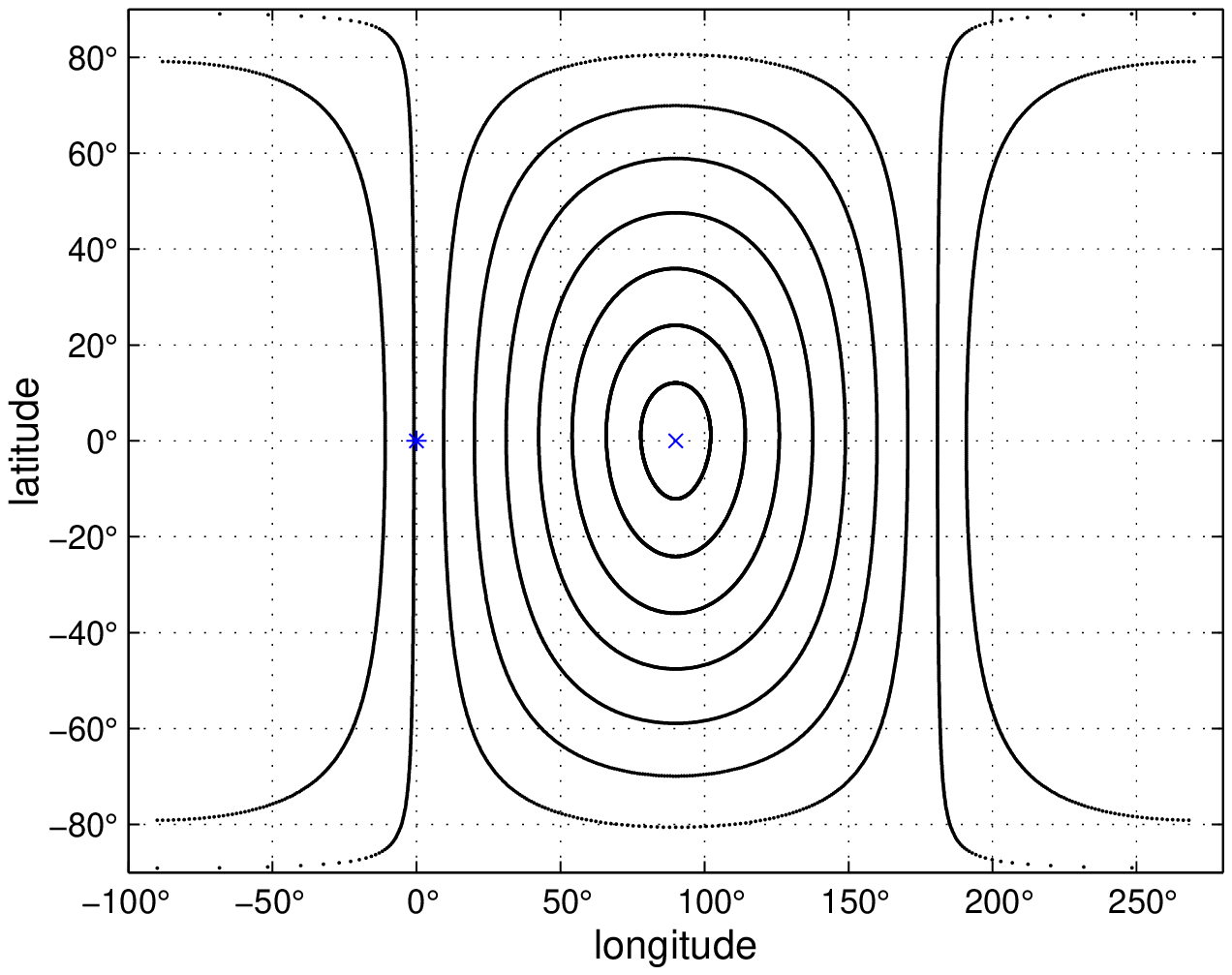,width=0.5\linewidth,clip=} &
\epsfig{file=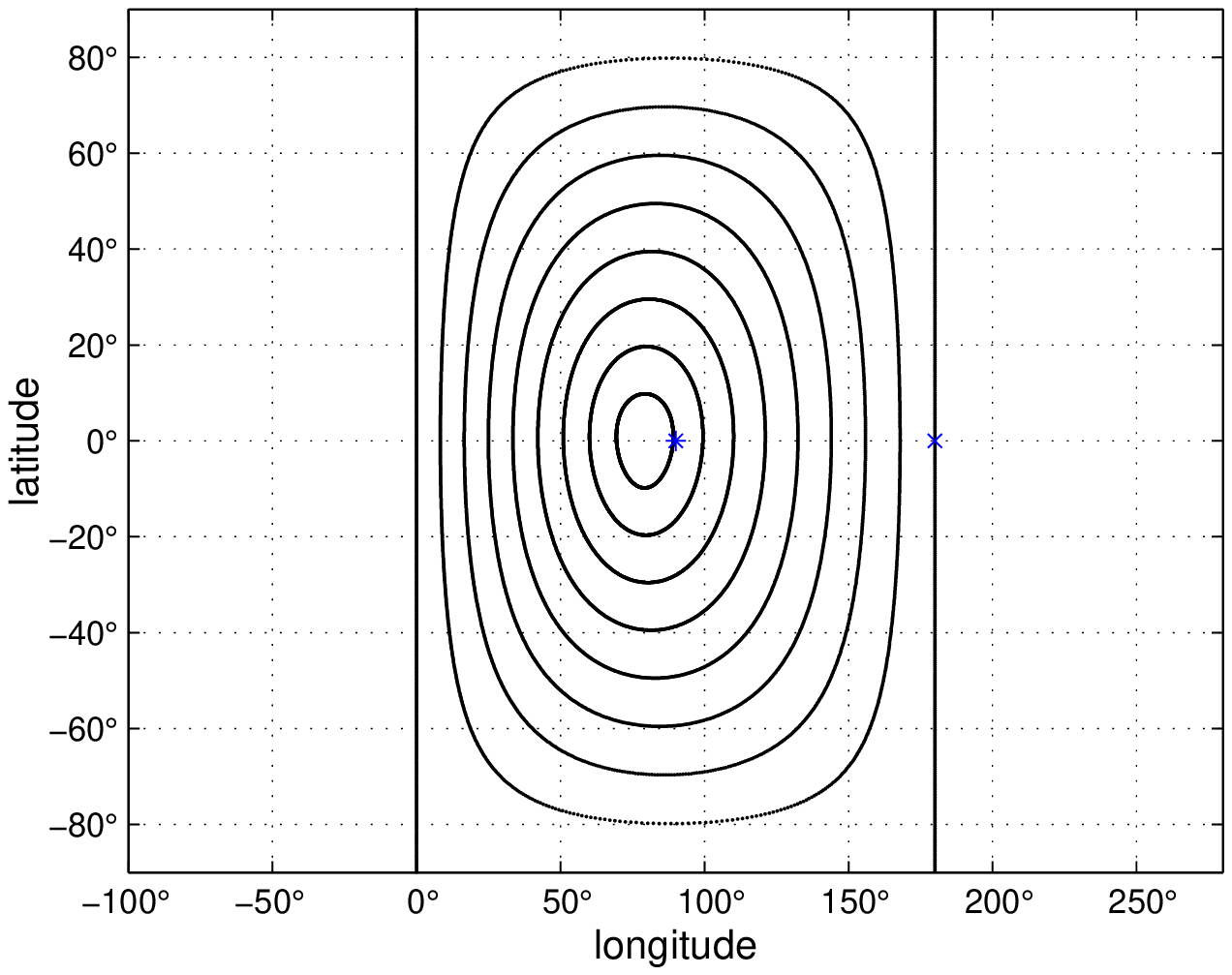,width=0.5\linewidth,clip=} \\
\epsfig{file=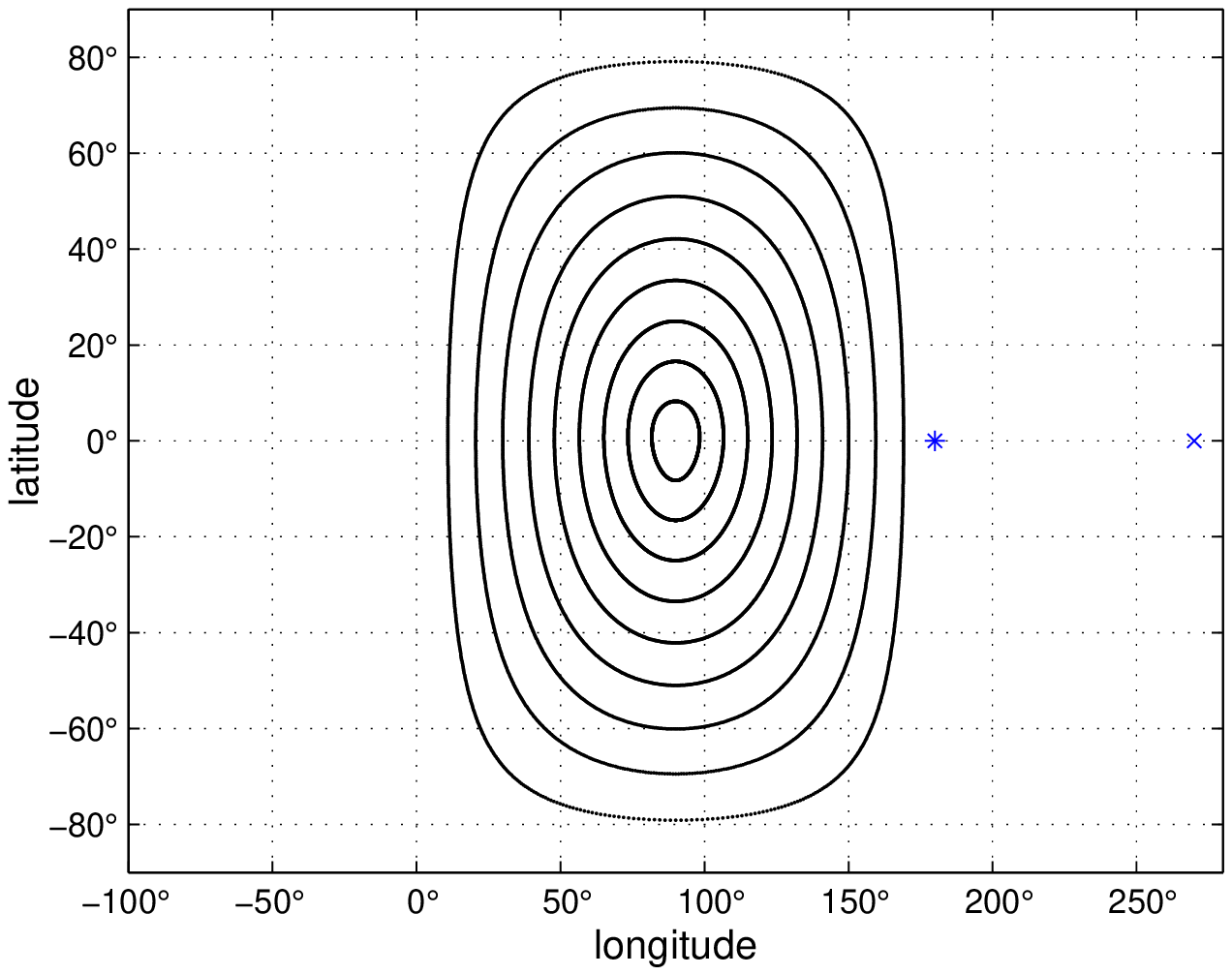,width=0.5\linewidth,clip=} &
\epsfig{file=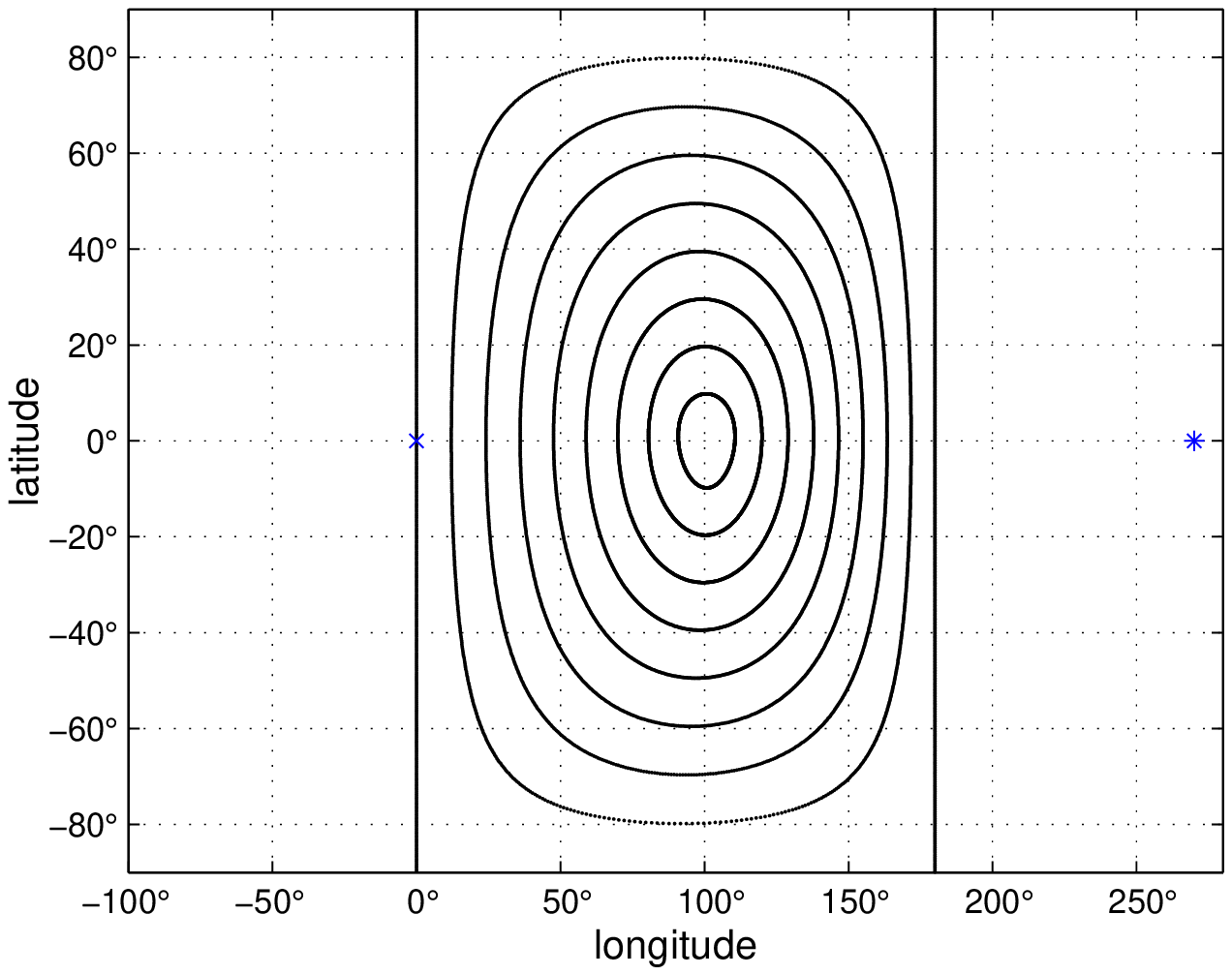,width=0.5\linewidth,clip=}
\end{tabular}
\end{center}
\caption{The distorted emission annulus for $r_{0}=30M$ and
$\lambda=0^{\circ}$. Each panel corresponds to a SMBH position
in Fig.~\ref{fig:sphere2} at $\varphi=0^{\circ}$ (upper
  left); $\varphi=90^{\circ}$ (upper right); $\varphi=180^{\circ}$ (lower
  left); $\varphi=270^{\circ}$ (lower right). In each panel, an asterisk at
zero latitude marks the SMBH position and a cross marks the direction
of the velocity of the SMBH relative to the pulsar.}
\label{fig:aberration2}
\end{figure}

We can now plot closed curves of emission directions for specific
choices of parameters.  By showing curves for values of $\alpha$ that
differ by $10^\circ$ we can identify the area between two closed curves 
as the beam annulus for many choices of $\alpha_0$ and $\Delta\alpha$.

Figures~\ref{fig:aberration1} and \ref{fig:aberration2} show curves
for $r_{0}=30M$. An observer fixed in position in the astronomical
reference frame will measure the pulsar to be moving at speed
$v/c=\sqrt{(M/r_0)/(1-2M/r_0)\;}=0.18898...$, the speed that is used 
in our Lorentz transformation. Figure~\ref{fig:aberration1} shows results for
$\lambda=45^\circ$, and Fig.~\ref{fig:aberration2} for
$\lambda=0^\circ$.  In each panel curves are given for 9 values of
$\alpha$ from $10^\circ$ (innermost) to $90^\circ$ (outermost), in $10^\circ$ increments; the
area between any two curves in this sequence can be taken as the beam
annulus. The 4 different panels in Fig.~\ref{fig:aberration1} and the
4 in Fig.~\ref{fig:aberration2} correspond to 4 different values of
$\varphi$, the location of the SMBH in its motion around the
pulsar (i.e., of the pulsar around the black hole). The value of
$\varphi$ in each panel is indicated by a small asterisk at zero
latitude.

For the most part, the distorted shapes in Figs.~\ref{fig:aberration1}
and \ref{fig:aberration2} are the manifestations of geometry, the
plotting of circles on a sphere in a cartesian coordinate system. The
relativistic aspects of these figures, i.e., those due to motion and
aberration, play two roles: (i)~They are the reason that there are
difference among the 4 panels in Figs.~\ref{fig:aberration1}, and 
that there are differences among the 4 panels in
Figs.~\ref{fig:aberration2}. (ii)~Aberration accounts for the small
left-right asymmetry that is apparent in some of the panels.  Both
these effects are small, because aberration effects are of order $v/c$, 
which is only around 0.19 for annuli shown in the figures.


\section{A specific scenario}\label{sec:scenarios}

The deflection of a pulsar beam towards an Earth-based telescope
requires an overlap.  The keyhole in Sec.~\ref{sec:keyholes}
describes the direction in which a photon must be emitted if it is to
be deflected toward the Earth. The distorted annulus in
Sec.~\ref{sec:annulus} describes the set of direction in which the
pulsar beam is deflected.  For there to be an observable event, the
keyhole direction must fall within the annulus. 

Determining when the keyhole enters and exits the annulus is
complicated because both the keyhole direction and the shape and
location of the annulus change with time, i.e., change as the pulsar
goes around the SMBH (or -- in our coordinate systems -- as the SMBH
goes around the pulsar). 

\begin{figure}[p]
\begin{center}
\includegraphics[width=.7\textwidth]{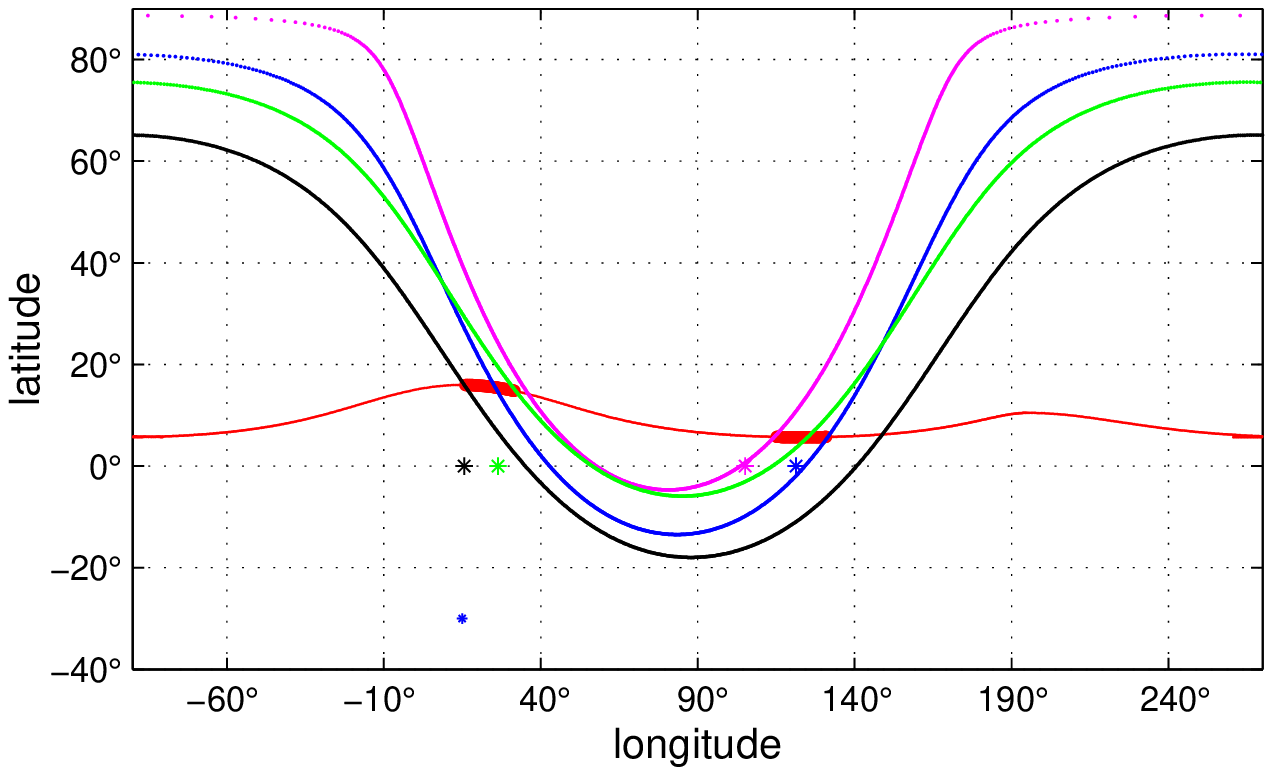}  
\caption{Events for a circular pulsar orbit with $r_0=30M$,
  $\lambda=45^\circ$, and a beam cone extending from $\alpha=50^\circ$
  to $\alpha=60^\circ$.
The axes are the longitude and latitude in the global frame
  in which the SMBH is stationary. 
The thin red line is the trajectory of
  the keyhole.
 } \label{fig:trial6}
\end{center}
\end{figure}

The formalism developed above makes no assumptions about the orbit of
the pulsar (it need not even be closed), and no assumptions about the
precession of the pulsar spin axis. The formalism can be applied to
any orbit, and any time variability of pulsar spin etc. In this
section, to illustrate the application of the formalism as clearly as
possible, we choose the pulsar to have a circular orbit, at radius
$r_0=30M$, and we ignore precession of the spin axis.  As in previous
sections we describe the motion as the SMBH going around the pulsar.
For our example, in Fig.~\ref{fig:trial6}, we take
$\lambda=45^{\circ}$, and take the pulsar beam to be confined between
$\alpha=50^\circ$ and $\alpha=60^\circ$. As pictured in
Fig.~\ref{fig:sphere2} the SMBH is moving counter-clockwise, and we
take it to be at $\varphi=90^\circ$ (i.e., in the $yz$ plane) at time
$t=0$. We choose to place the Earth observer at longitude
$\varphi_{\oplus}=15^{\circ}$ and latitude
$\lambda_{\oplus}=-30^{\circ}$. The Earth observer location is shown
as a small blue asterisk in Fig.~\ref{fig:trial6}.

In Fig.~\ref{fig:trial6}, the thin red curve represents the trajectory
of the keyhole. Because we have chosen to have the SMBH start at
$\varphi=90^\circ$, the corresponding keyhole starts at around
$100^\circ$ and is moving in the direction of increasing longitude
$\varphi$.  When the SMBH moves to the longitude marked by the
magenta asterisk, at $\varphi=105.16^\circ$, the keyhole moves into 
the annulus; the thin red curve of keyhole
location crosses the magenta curve which represents the
$\alpha=50^\circ$ boundary of the beam annulus at that moment.  At
this crossing event, pulses start to become detectable by the Earth
observer. This period of observability (thickened segment of the keyhole
trajectory) extends to the configuration at
which the SMBH is at $\varphi=121.31^\circ$, denoted by the large
blue asterisk, the SMBH position that corresponds to the trajectory of the keyhole
(the thin red curve) passing the blue curve representing the 
$\alpha=60^\circ$ boundary of the beam annulus at that moment of crossing.
The number of pulses observable depends on the spin rate of the pulsar,
but in any case will be large since the pulsar spin period is 8 or more
orders of magnitude smaller than the orbital period.

After more than half an orbit passes, the SMBH is at
$15.63^\circ$ (black asterisk) and the keyhole trajectory passes the
black curve representing the $\alpha=60^\circ$ boundary of the beam
annulus. This starts a period of pulse observability which ends when
the SMBH is at $26.43^\circ$ (green asterisk), corresponding to the keyhole
trajectory passing the green curve that represents the
$\alpha=50^\circ$ boundary at that crossing.

It should be understood that the green and the magenta curves both
represent the directions of photons emitted at $\alpha=50^\circ$, but
they represent those directions for different locations of the SMBH
relative to the pulsar.  Similarly, the black and blue curves
both represent the directions of photons emitted at $\alpha=60^\circ$.

\begin{figure}[h]
\vspace{-.25in}
\begin{center}
\begin{tabular}{cc}
\epsfig{file=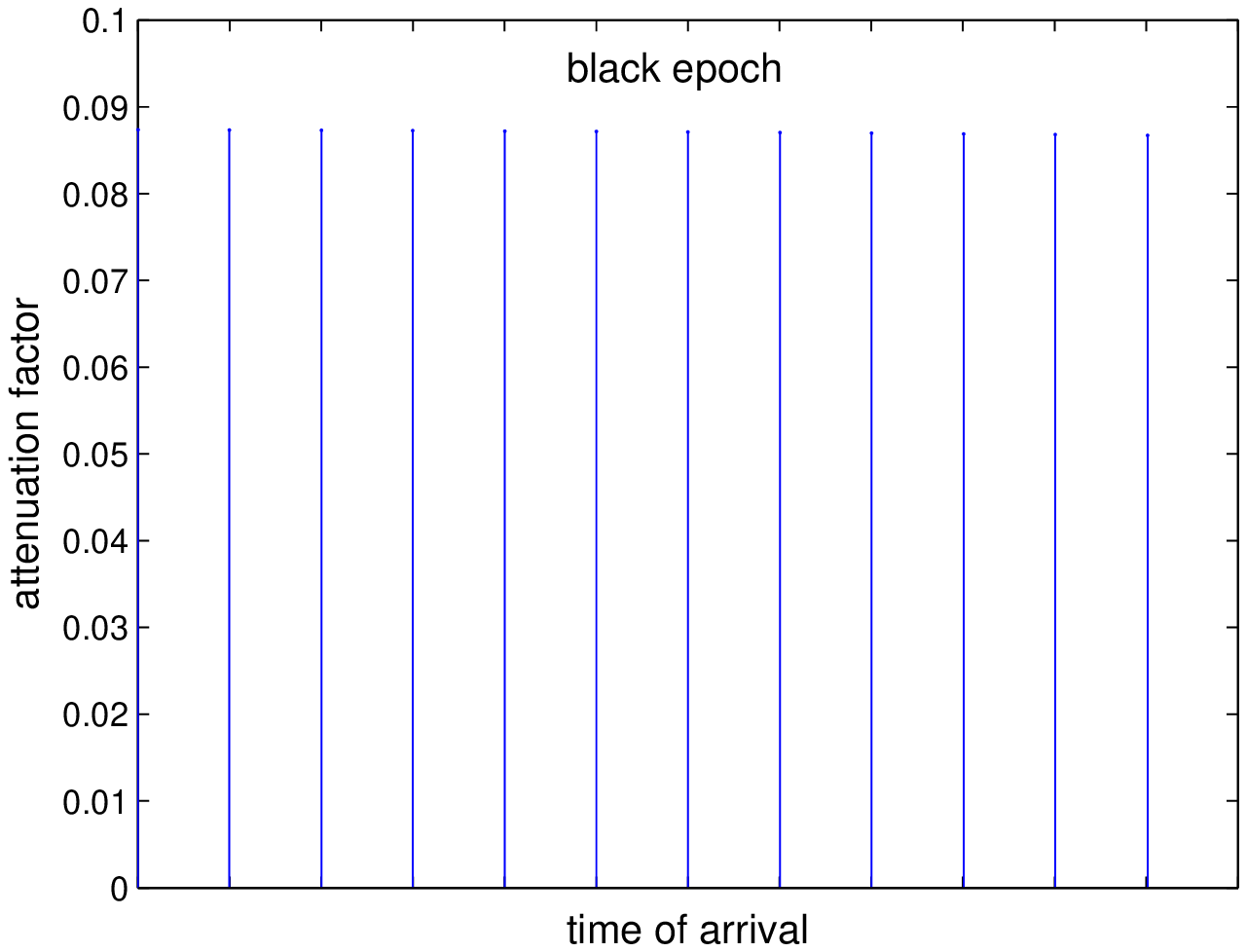,width=0.45\linewidth,clip=} &   
\epsfig{file=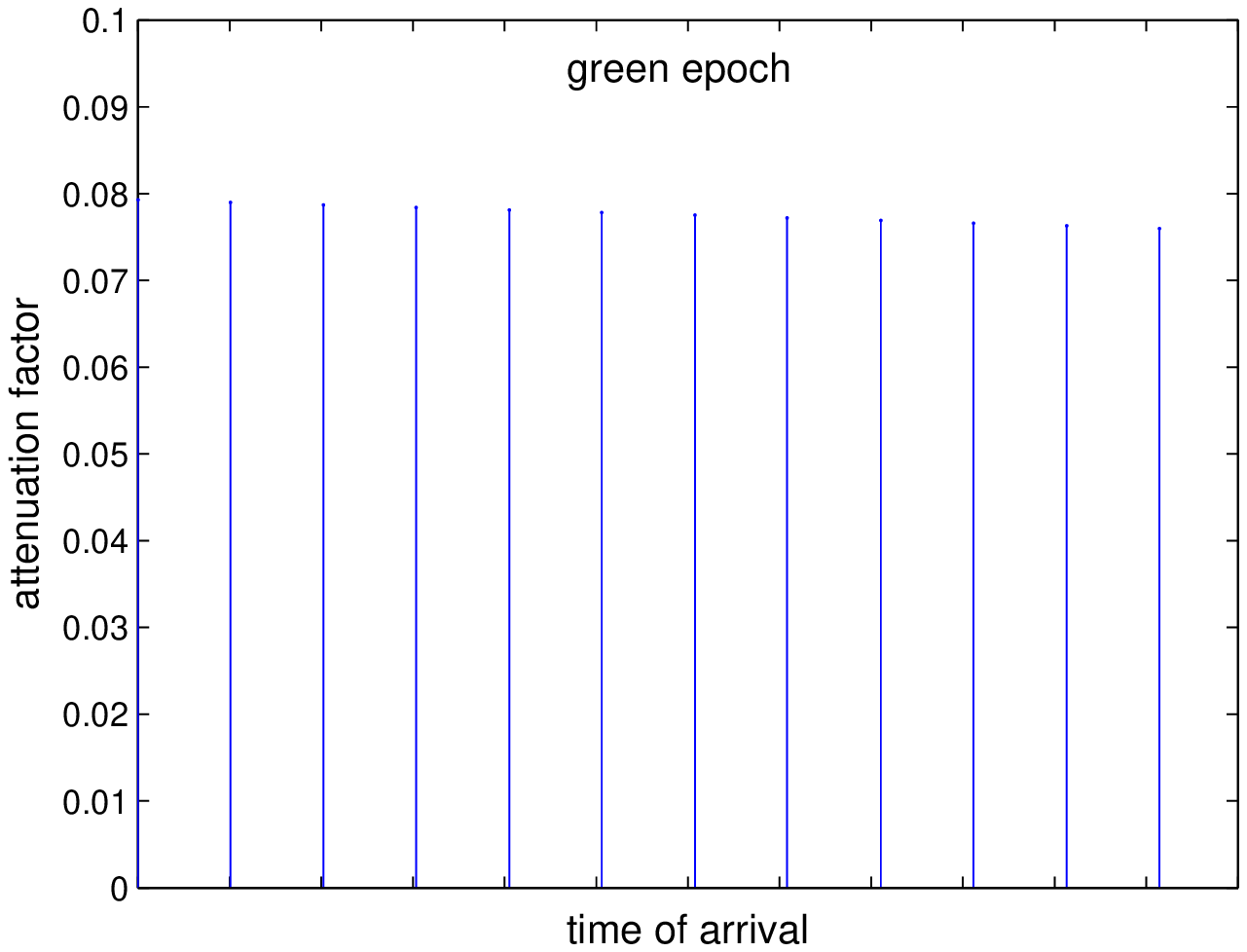,width=0.45\linewidth,clip=} \\
\epsfig{file=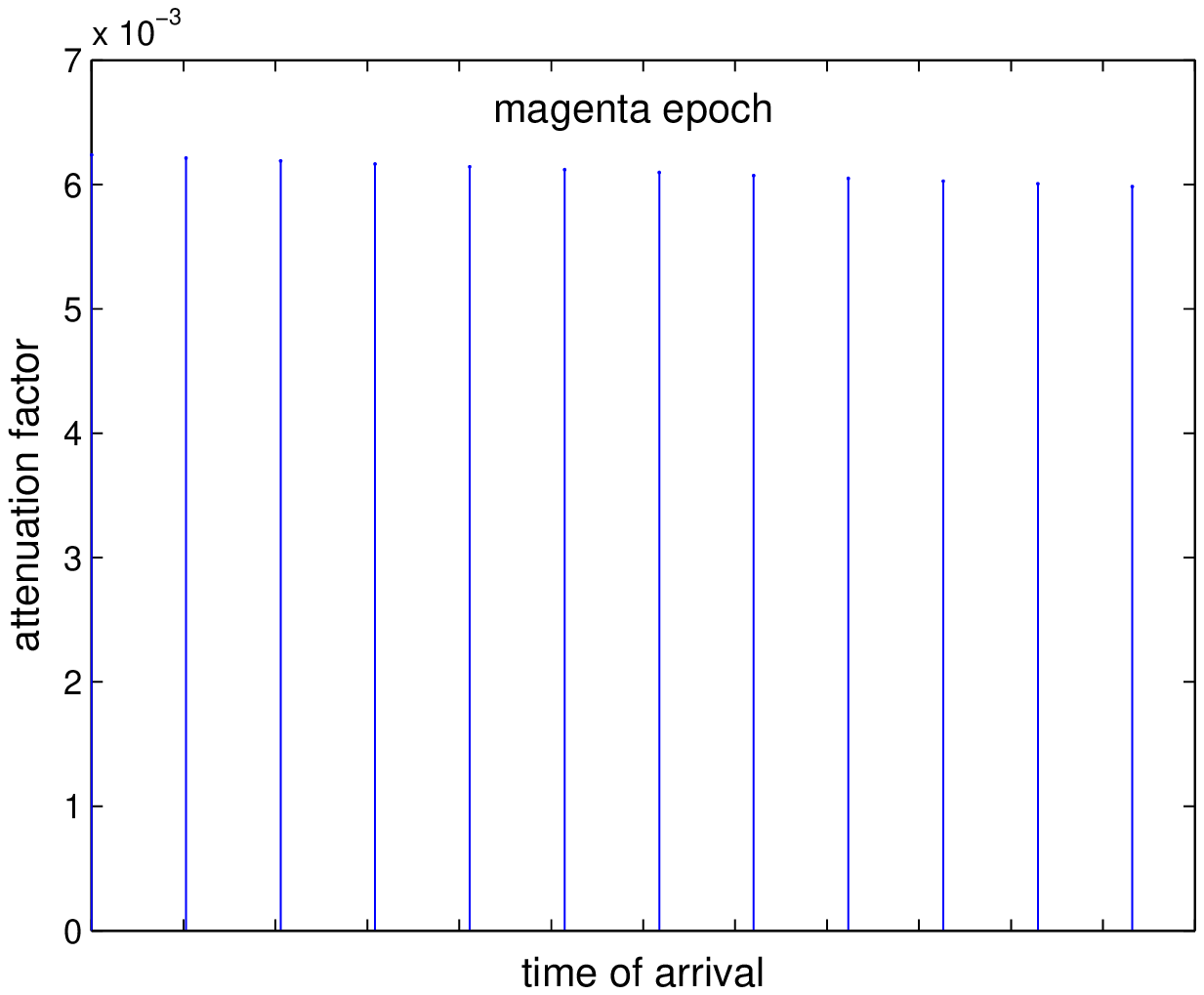,width=0.45\linewidth,clip=} &
\epsfig{file=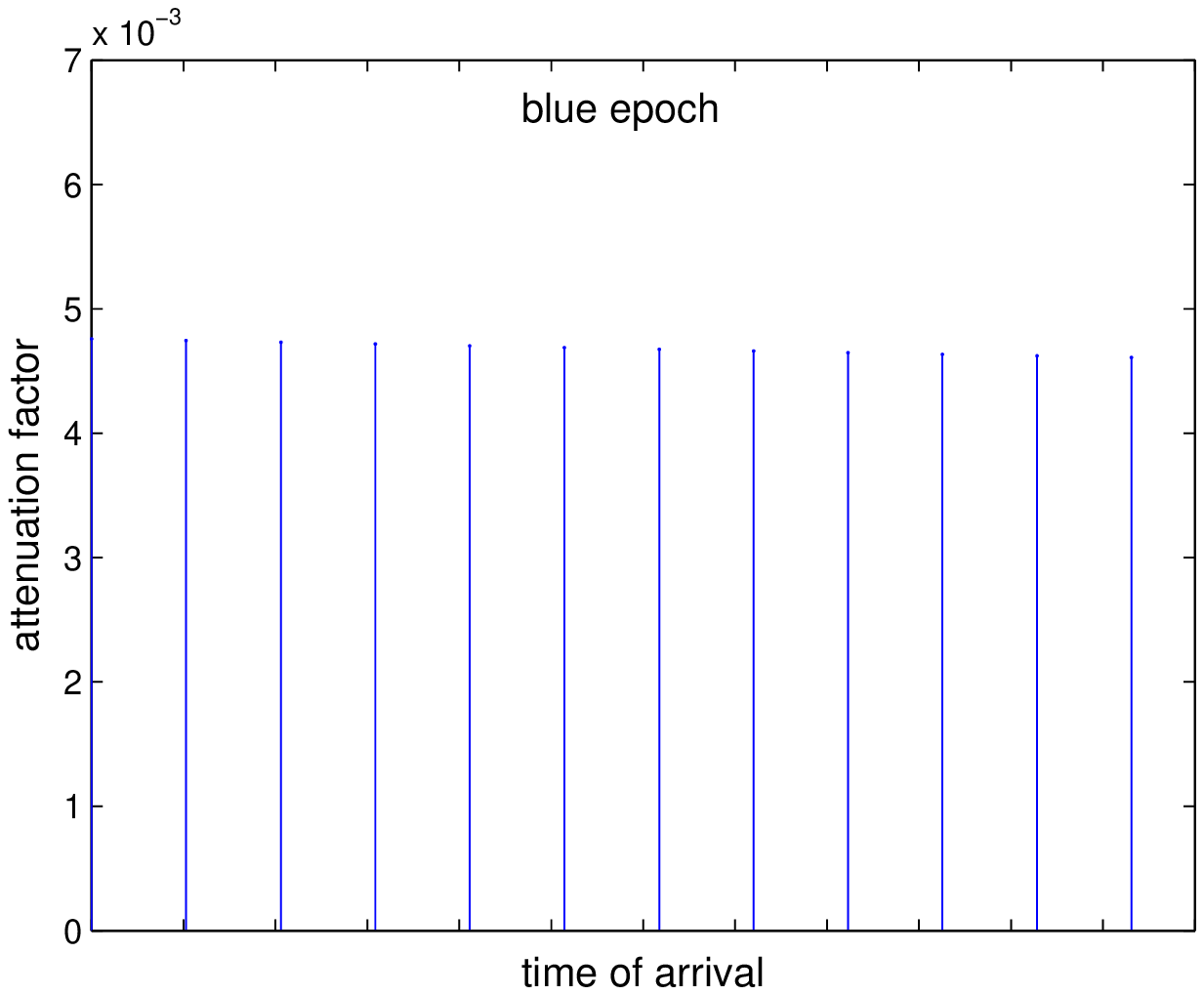,width=0.45\linewidth,clip=}
\end{tabular}
\end{center}
\caption{In each panel above, the horizontal axis is the time of
arrival of pulses observable by the Earth observer, and the vertical
axis is the amplitude factor of Eq.~\eqref{Amp}. See text for details. 
}
\label{fig:ABCD}
\end{figure}

In Fig.~\ref{fig:ABCD}, we represent sequences of pulses for four
epochs: black (corresponding to a time shortly after the passage of
the keyhole trajectory crosses the black annulus curve), green,
magenta and blue (similarly).  The unit of time in each panel is the
``proper'' pulsar spin period, the spin period measured by an observer
comoving with the pulsar. The value of that unit, in terms of $M$, or
of seconds, need not be specified.  We assume only that the unit is
many orders of magnitude less than the pulsar orbital period, so that
many pulses are emitted with no substantial change in the
pulsar-SMBH-Earth configuration. Without specifying a particular 
pulse period, we can compare the period of reception of pulses
characteristic of each epoch.

The two upper panels show several pulses (vertical lines) from the
epochs of the black and green keyhole-annulus crossings.  The
alignment of the pulses with the timing marks in Fig.~\ref{fig:ABCD}
shows that the observed pulse period, in both cases, is slightly
longer than the proper pulse period, although it is barely noticeable
in the case for the black epoch). For both events the pulsar is moving
almost transverse to the direction to the Earth. The photon path from 
the pulsar to the Earth is, therefore, only very slightly affected by
the pulsar motion, and the effect on the received pulse period 
is negligible.
The lower two panels correspond to the magenta and blue
epochs. In both cases the pulsar is moving towards the Earth at pulse
emission. For these epochs the photon path starts out generally away
from the Earth, winds around the SMBH and proceeds towards the Earth.
As the pulsar moves toward the Earth this backward then forward photon
path increases in length thereby producing a redshift-like lengthening
of the period between pulses. This pulse-period lengthening gives a
noticeable difference between the observed and the proper pulse
period.

The effects of attenuation, described by Eq.~\eqref{Amp} are even more
noticeable. For the upper panels showing the epochs of the black and
the green events, the intensity is approximately 9\% and 8\%
respectively of the intensity of a direct beam. For the lower panels
showing the epochs of the magenta and blue events the attenuation is
greater; the beam intensity is approximately 0.6\% and 0.5\% that of a
direct beam.  These attuations are not due to photon redshifts. For these epochs
the pulsar is not receding from  the Earth and, in any case, the red/blueshift
effects are very small and are not included in Eq.~\eqref{Amp}, for
reasons explained in Sec.~\ref{sec:keyholes} and in Paper I.


\section{Summary and Conclusions}\label{sec:conc}

We have set down the principles and the methods of calculation for
analyzing the observational effects of a nonspinning black hole on a
pulsar signals.  In particular, we have shown that the calculation of
pulse timing and intensity involves solving two problems
simultaneously: that of the the beam annulus distorted by aberration,
and that of the keyhole position and trajectory. Once these calculations
are done, one can quickly infer the strong field effect on the pulse
period and the beam attenuation for any epoch of motion. 

The beam annulus computation starts with the pulsar's beam details:
the direction of its spin axis, and the orientation and width of its
radio beam. Photon directions, most naturally specified in the frame
of the pulsar, must then be transformed to the astronomical frame so 
that aberration effects are taken into account. The algorithm for 
doing this is set down in Sec.~\ref{sec:annulus}.

For any pulsar-SMBH-Earth alignment, the keyhole direction is the
photon emission direction (in the astronomical frame) for which a
photon will be deflected towards the Earth. This direction must lie in
the pulsar-SMBH-Earth plane, and the necessary deflection angle is
determined by the pulsar-SMBH-Earth angles in that plane. The keyhole
direction in that plane is inferred from the deflection angle by
inverting the universal function $F$ developed in Paper I.  The
specific calculational steps are given in Eqs.~\eqref{algorithm1}--
\eqref{algorithm4}.  The relative simplicity of this method is the
main motivation for ignoring the rotation of the SMBH.

Once the pulsar trajectory is specified, the two parts of the analysis
are brought together.  For any pulsar-SMBH-Earth alignment, it is
determined whether the keyhole overlaps the beam annulus. Since the
beam annulus shape, due to aberration effects, depends on the pulsar
velocity, the shape must be calculated for each new position of the
pulsar.

Though the use of the universal function greatly simplifies the
calculations, the exploration of the parameter space for pulsars in
Sgr A$^*$ is still inconveniently broad.  This is particularly true
because pulsar orbits near Srg A$^*$ may be highly eccentric, thereby
adding to the already large set of parameters that constitute a model.

A minor but useful simplification is to ignore the aberration effects
on the beam annulus, so that the beam annulus is fixed in the
astronomical frame once and for all. (Precession of the pulsar spin
axis is also ignored.) The aberration, in any case, is of order of the
pulsar velocity divided by $c$, and will be small.  We have seen that
the effects are small with the choice we have made, $r_0=30M$ for the
model in Secs.~\ref{sec:annulus} and \ref{sec:scenarios}.  Yet, even
for orbits of high eccentricity, $r_0=30M$, a velocity $v/c\approx0.19$
is much more relativistic than what is likely to be the case for
pulsars that are beaming past the SMBH. 

A more detailed discussion of plausible astrophysical models will be
given elsewhere \citep{probpaper}. Here we only point out that an
epoch of observability will typically be of order of 10$^\circ$ or
more of the pulsar orbit.  This translates to millions of seconds for
the most tightly bound pulsars in the Galactic center. Since pulsar
spin periods are typically less than, or much less than a second, the
pulse train for an epoch of observability will involve a large number
of pulses. The timing and intensity patterns of that pulse train can
reveal much about the strong field region through which the pulses
traveled.

\section{Acknowledgment} We gratefully acknowledge support by the
National Science Foundation under grants AST0545837, PHY0554367,
and 0734800. We
also thank the NASA Center for Gravitational Wave Astronomy at
University of Texas at Brownsville. YW acknowledges support by the
Chinese National Science Foundation under grant 10773005.


\begin{thebibliography}{12}
\expandafter\ifx\csname natexlab\endcsname\relax\def\natexlab#1{#1}\fi

\bibitem[{{Campana} {et~al.}(1995){Campana}, {Parodi}, \&
  {Stella}}]{campanaparodistella95}
{Campana}, S., {Parodi}, A., \& {Stella}, L. 1995, \mnras, 277, 1162

\bibitem[{{Creighton} {et~al.}(2009){Creighton}, {Price}, {Wang}, \&
  {Jenet}}]{probpaper}
{Creighton}, T., {Price}, R.~H., {Wang}, Y., \& {Jenet}, F.~A. 2009, in
  preparation

\bibitem[{{Freitag} {et~al.}(2006){Freitag}, {Amaro-Seoane}, \&
  {Kalogera}}]{freitagetal2006}
{Freitag}, M., {Amaro-Seoane}, P., \& {Kalogera}, V. 2006, \apj, 649, 91

\bibitem[{{Goicoechea} {et~al.}(1992){Goicoechea}, {Mediavilla}, {Buitrago}, \&
  {Atrio}}]{goicoecheaetal92}
{Goicoechea}, L.~J., {Mediavilla}, E., {Buitrago}, J., \& {Atrio}, F. 1992,
  \mnras, 259, 281

\bibitem[{{Gorham}(1986)}]{gorham86}
{Gorham}, P.~W. 1986, \apj, 303, 601

\bibitem[{{Laguna} \& {Wolszczan}(1997)}]{lagunawolszcan97}
{Laguna}, P., \& {Wolszczan}, A. 1997, \apjl, 486, L27

\bibitem[{{Lazio} {et~al.}(2006){Lazio}, {Deneva}, {Bower}, {Cordes}, {Hyman},
  {Backer}, {Bhat}, {Chatterjee}, {Demorest}, {Ransom}, \&
  {Vlemmings}}]{lazioetal2006}
{Lazio}, J., {Deneva}, J.~S., {Bower}, G.~C., {Cordes}, J.~M., {Hyman}, S.~D.,
  {Backer}, D.~C., {Bhat}, R., {Chatterjee}, S., {Demorest}, P., {Ransom},
  S.~M., \& {Vlemmings}, W. 2006, Journal of Physics Conference Series, 54, 110

\bibitem[{{Melia}(2007)}]{meliabook}
{Melia}, F. 2007, {The Galactic Supermassive Black Hole} (Princeton: Princeton
  University Press)

\bibitem[{{Muno} {et~al.}(2008){Muno}, {Baganoff}, {Brandt}, {Morris}, \&
  {Starck}}]{munoetal2008}
{Muno}, M.~P., {Baganoff}, F.~K., {Brandt}, W.~N., {Morris}, M.~R., \&
  {Starck}, J.-L. 2008, \apj, 673, 251

\bibitem[{{Oscoz} {et~al.}(1997){Oscoz}, {Goicoechea}, {Mediavilla}, \&
  {Buitrago}}]{oscozetal97}
{Oscoz}, A., {Goicoechea}, L.~J., {Mediavilla}, E., \& {Buitrago}, J. 1997,
  \mnras, 285, 413

\bibitem[{{Pfahl} \& {Loeb}(2004)}]{pfahlloeb04}
{Pfahl}, E., \& {Loeb}, A. 2004, \apj, 615, 253

\bibitem[{{Wang} {et~al.}(2009){Wang}, {Jenet}, {Creighton}, \&
  {Price}}]{paperI}
{Wang}, Y., {Jenet}, F.~A., {Creighton}, T., \& {Price}, R.~H. 2009, \apj, 697,
  237, (Paper I)

\end{thebibliography}
\end{document}